\documentclass[aps,prd,reprint,superscriptaddress,nofootinbib,showpacs,twoside,notitlepage]{revtex4-1}    
\usepackage{graphicx}
\usepackage{euscript,amssymb}
\usepackage{amsfonts}
\usepackage{amssymb}
\usepackage{amsbsy}
\usepackage{amsmath}
\usepackage{nccmath}   
\usepackage[dvips]{color}
\usepackage{color}
\usepackage{bbold}

\newcommand{\be}{\begin{equation}}
  \newcommand{\ee}{\end{equation}}
\newcommand{\ben}{\begin{eqnarray*}}
  \newcommand{\een}{\end{eqnarray*}}
\newcommand{\bea}{\begin{eqnarray}}
  \newcommand{\eea}{\end{eqnarray}}
\newcommand{\bdm}{\begin{displaymath}}
  \newcommand{\edm}{\end{displaymath}}
\newcommand{\ba}{\begin{align}}
  \newcommand{\ea}{\end{align}}

\newcommand{\del}{\partial}

\newcommand{\intx}{\int\!\mathrm{d}^3x}
\newcommand{\inttx}{\int\!\mathrm{d}t\,\mathrm{d}^3x\,}
\newcommand{\intxx}{\int\!\mathrm{d}^4x\,}

\newcommand{\ssst}{\scriptscriptstyle}

\begin{document}

\title{Conformal and Weyl-Einstein gravity: Classical geometrodynamics}

\author{Claus Kiefer}

\email{kiefer@thp.uni-koeln.de}

\author{Branislav Nikoli\'c}

\email{nikolic@thp.uni-koeln.de}

\affiliation{Institut f\"ur Theoretische Physik, Universit\"{a}t zu
K\"{o}ln, Z\"{u}lpicher Stra\ss e 77, 50937 K\"{o}ln, Germany}

\date{\today}

\begin{abstract}
We present a new formulation for the canonical approach to conformal
(Weyl-squared) gravity and its extension by the Einstein-Hilbert term
and a nonminimally coupled scalar field. 
For this purpose we use a unimodular decomposition of the three-metric
and introduce unimodular-conformal canonical variables. The important
feature of this choice is that only the scale part of the three-metric
and the rescaled trace part of the extrinsic curvature change under a
conformal transformation. This significantly simplifies the constraint
analysis and manifestly reveals the conformal properties of a theory
that contains the conformally invariant Weyl-tensor term. The
conformal symmetry breaking which occurs in the presence of the
Einstein-Hilbert term and a nonconformally coupled scalar field can
then be interpreted directly in terms of this scale and this trace. We
also discuss in detail the generator for the conformal
transformations. This new Hamiltonian formulation is especially
suitable for quantization, which will be the subject of a separate
paper. 
\end{abstract}

\maketitle


\section{Introduction}
\label{Intro}

Gravitational theories beyond general relativity (GR) are addressed
for various reasons. One is the conceptual need to accommodate gravity
into the quantum framework \cite{OUP}. Another is the attempt to
describe cosmological features, notably Dark Matter and Dark Energy,
by generalized classical theories \cite{CF11}. In this paper, we deal
with conformal (Weyl-squared) gravity  
(also called W theory below) both as a pure gravitational theory and as
part of an action containing also an Einstein-Hilbert (EH) part and an
action describing a (in general nonminimally coupled) scalar
field. Our main reason for doing so is quantum gravity. Researchers
have often entertained the idea that at a fundamental level, for
example at high energy, Nature can be described by a scale free theory,
with scales emerging only at lower energy (see
e.g. \cite{Hooft15}). In order to study the consequences of a
scale free theory we investigate here conformal (Weyl-squared) gravity
described by an action containing the square of the Weyl
tensor. Historically, this has emerged as an offspring from Weyl's
original gauge theory published in 1918; see for example \cite{KN17a}
for a brief historical review. Conformal gravity was later used (and
still is today) as an alternative classical theory to GR with
potential astrophysical implications \cite{Mannh2012} and as an
emerging contribution to the effective gravitational action at the
one-loop level of quantum field theory in curved spacetime
\cite{ParkerToms,BOS}.   
These are not the aspects we are interested in here. We take conformal
gravity as a {\em model} for a conformally invariant theory whose
quantum version might be relevant at the most fundamental level.  

An approach especially suited for conceptual questions and for
cosmological applications is the canonical (Hamiltonian) approach
\cite{OUP}. This is the subject of this paper.  
We take it as a preparation for the quantum theory discussed in a
forthcoming contribution, but find the classical discussion
interesting in its own right because, as we shall see, interesting
conceptual and mathematical structures appear there. These will mainly
concern the structure of the constraints and the generator for
conformal transformations.  

The Hamiltonian formalism of conformal gravity has already been the
subject of various investigations. Important earlier contributions
include \cite{Kaku1982,Blw,Hwz}, and \cite{BOS}.  
In \cite{DerrHD}, the Hamiltonian formalism of $f({\rm Riemann})$
theories as well as of Weyl gravity was studied. More recently, the
authors of \cite{Kluson2014} extended the analysis to the Hamiltonian
formulation of the Weyl action plus conformally coupled scalar fields as
well as their extension by including the Einstein-Hilbert term and an
$R^2$ term; they analyzed, in particular, the constraint algebra in
great detail. The authors of \cite{ILP} investigated a
particular model within Weyl-squared gravity, with an important step being the
derivation of the generator of gauge conformal transformations using
the Castellani algorithm \cite{Cast}.  

In our paper here, we shall develop a new version of the Hamiltonian
formalism, which is especially suited for quantization. We shall use
an irreducible decomposition of the three-metric into its scale
(determinant) part and its conformally invariant (unimodular) part. In
this way, new canonical variables will be identified. A similar
procedure will be applied to the lapse function and the shift vector,
leading to densities. This will then {\em induce} a decomposition of
the extrinsic curvature into its {\em rescaled} traceless and trace
parts.\footnote{In some of the earlier work, e.g. \cite{Kaku1982} and
  \cite{Blw}, the extrinsic curvature was decomposed in this way by
  hand, but the unimodular decomposition was not used.} The resulting
variables are all densities and are all conformally invariant except
the scale (determinant) and the rescaled trace of the extrinsic
curvature. One may refer to this approach as the {\em Hamiltonian
  formalism in unimodular-conformal variables}. It will turn out that
the use of these new variables reveals the full power of the conformal
invariance associated with the Weyl-tensor part of the full action and
simplifies the discussion of conformal symmetry breaking induced by
other terms.  

Our paper is organized as follows. 
In Sec.~II, we present a brief review of the Weyl-tensor action and
some of its implications. Section~III contains a summary of the $3+1$
decomposition for Weyl-squared gravity. Our own contributions start with
Sec.~IV. There, we introduce the unimodular-conformal variables in
configuration space, thus dividing the variables into a conformally
invariant and a noninvariant sector. Section~V  treats pure Weyl
gravity. It is divided into four parts. Part~A is devoted to the
Hamiltonian formalism and the 
constraint analysis. Part~B deals with the constraint algebra and
part~C with the generator of conformal transformations. Part~D
presents the Hamilton-Jacobi functional. Section~VI is devoted to Weyl
gravity plus the EH term. It is divided into two parts. In part~A, we
present the Hamiltonian 
formulation and in part~B the Hamilton-Jacobi functional and nongauge
transformations. The addition of the EH term explicitly leads to
conformal symmetry breaking, for which only two variables containing a
scale are responsible: the determinant of the three-metric and the
trace of the extrinsic curvature. In that section we shall also argue
that, in spite of some of the constraints being second class, it is
still possible to define the generator of conformal but {\em nongauge}
transformations.  
 In Sec.~VII, a nonminimally coupled scalar field is added, with the
 Hamiltonian formalism presented in part~A and the Hamilton-Jacobi
 functional together with the generator of conformal transformations
 presented in part~B. Section~VIII contains our conclusions. 
 We also have some Appendices. The first
 Appendix contains a short discussion of the physical dimensions; the
 remaining three Appendices present technical details for the
 discussion in the body of our paper.

\section{Preliminaries}
\label{Prelim}

In this section, we shall present a short summary of the covariant formulation for
conformal gravity \cite{Kaku1982,Blw,Mannh2012,Kluson2014}. This
theory is defined by the action 
\begin{equation}
\label{eqn:W-action}
S^{\ssst \rm W}:=-\frac{\alpha_{\ssst\rm W}}{4}\int{\rm
  d}^4x\sqrt{-g}\,C_{\mu\nu\lambda\rho} 
C^{\mu\nu\lambda\rho}\,,
\end{equation}
where $\alpha_{\ssst\rm W}$ is a coupling constant with the dimension
of an action, and
\begin{equation}
\label{eqn:Weyltensor}
 {C^{\mu}}_{\nu\lambda\rho}={R^{\mu}}_{\nu\lambda\rho}-\left(\delta^{\mu}_{[\lambda}R_{\rho]\nu}-g_{\nu[\lambda}{R^{\mu}}_{\rho]}\right)-\frac{1}{3}\delta_{[\rho}^{\mu}g_{\lambda]\nu}R
\end{equation}
is the Weyl tensor, which is invariant under conformal transformations of the metric,
\begin{equation}
\label{eqn:cftrans}
g_{\mu\nu}(x)\quad\rightarrow\quad \tilde{g}_{\mu\nu}(x)=\Omega^{2}(x)g_{\mu\nu}(x).
\end{equation}
We refer to this model simply as ``Weyl-squared theory'' or
``W theory''. The field equations following from \eqref{eqn:W-action}
read  
\begin{equation}
\label{eqn:Bach}
\left(\nabla_{\mu}\nabla_{\nu}+\frac{1}{2}R_{\mu\nu}\right){C^{\mu}}_{\lambda\,\,\,\rho}^{\,\,\,\nu}=0.
\end{equation}
In order to study the breaking of conformal symmetry, we shall also investigate below the extension of \eqref{eqn:W-action} by the EH term and the action for a nonminimally coupled scalar field $\varphi$,
\begin{align}
\label{eqn:WEHm-action}
S^{\ssst \rm WE\varphi}:=\intxx\sqrt{-g}\Biggl[ &-\frac{\alpha_{\ssst\rm W}}{4}\,C_{\mu\nu\lambda\rho}
C^{\mu\nu\lambda\rho}+\frac{1}{2\kappa}R\nonumber\\
&-\frac{1}{2}\biggl(g^{\mu\nu}\del_{\mu}\varphi\del_{\nu}\varphi+\xi R\varphi^2 \biggr)
\Biggr],
\end{align}
where $\kappa:= 8\pi G$, and $\xi$ is the dimensionless nonminimal
coupling constant. We will refer to the extended models as
``Weyl-Einstein'' (WE) and WE$\chi$ 
theories,\footnote{$\chi$ because below we shall introduce a rescaled
  field $\chi$.} respectively. Note that for 
 $\xi=1/6$, the scalar field is conformally invariant (see e.g. \cite{ParkerToms}, Sec. IIB).
We use units with $c=1$ throughout.

For a general field $\phi_{A}(x)$, the conformal transformation is implemented by
\begin{equation}
\label{eqn:cftransF}
\phi_{\ssst A}(x)\quad\rightarrow\quad \tilde{\phi}_{\ssst A}(x)=\Omega^{n_{A}}(x)\phi_{\ssst A}(x),
\end{equation}
where $A$ denotes a collection of spacetime and/or internal indices, $\Omega(x)$ is a positive function of the spacetime coordinates, and $n_{A}$ is a rational number that is characteristic for each field and called ``conformal weight''. More appropriately, this transformation is called \textit{local Weyl rescaling} or \textit{local dilatational transformation} because it is a
transformation of the fields themselves and not of coordinates. We are thus not talking about
the 15-parameter conformal group; see, for example, \cite{Kast,Nak,IOSW} for details.
 The transformation \eqref{eqn:cftrans} expresses the fact that the covariant metric tensor is of conformal weight 2. 
 
 Weyl gravity is an example of a theory with higher derivatives. For such theories, various subtleties occur. 
When linearizing the pure W theory, one finds that it contains a massless spin-2 state
(not yet the graviton), a massless spin-1 state, and also a massless spin-2 \textit{ghost} (negative-energy) state, adding up to six degrees of freedom in total as shown by Riegert \cite{Rieg}. The existence of ghost states is, in fact, not surprising.
In the canonical formalism, there is in the Hamiltonian a term linear
in the momentum, which signals that the energy is unbounded from
below. 
  This is usually called ``Ostrogradski instability''; see for example
  \cite{Wood}.\footnote{Also referred to as \textit{Theorem of
      Ostrogradski}, which states that any nondegenerate Lagrangian
    containing second or higher (but even-order) time derivatives
    gives rise to the existence of both positive and negative energy
    states. It has recently been shown that this conclusion can be
    extended to odd-order derivatives as well, including the case of a
    degenerate Lagrangian \cite{MotoSuy}.}  
  It can also be deduced from the corresponding propagator in a
  perturbative approach \cite{Stelle77}. It follows, in particular,
  that the W theory supplemented by the EH-term contains two degrees
  of freedom in one massless spin-2 propagator and five  
  degrees of freedom in one massive spin-2 \textit{negative energy} 
  propagator, amounting to seven in total. Negative energy states can
  be traded for positive ones, with the price of unitarity violation
  \cite{HH}. Which representation of Ostrogradski instability one
  takes depends on the context of the theory and the approach in
  question. The issue of Ostrogradski instabilities and ghosts has
  been addressed so far on a number of occasions, for example partial
  masslessness \cite{DJW}, critical gravity \cite{LuP}, and by
  introducing a PT-symmetric Hamiltonian (where P stands for parity
  reversal and T for time reversal) \cite{BendMan}. It is still an
  open problem and we do not solve it here either, but we shall
  reveal a new perspective by which it could be eventually solved.

\section{$3+1$ decomposition of Weyl-squared gravity}
\label{ADM}

We shall employ here a $3+1$ decomposition of spacetime, for which the
foliation is performed 
in terms of three-dimensional spacelike hypersurfaces $\Sigma_t$
parametrized by a time function $t$; 
see, for example, \cite{OUP} or \cite{Wald}. 
 A normalized covariant four-gradient of the time function is used to
 define a timelike unit covariant four-vector
 $n_{\mu}=-N\nabla_{\mu}t$, where $N>0$ is the lapse function, and we
 have the normalization $g_{\mu\nu}n^{\mu}n^{\nu}=-1$. In Arnowitt-Deser-Misner (ADM)
 variables, the covector $n_{\mu}$ has components
 $n_{\mu}=(-N,0,0,0)$, while its contravariant version is given by
 $n^{\mu}=(1/N,-N^{i}/N)$, where $N^{i}$ is the shift vector. The
 decomposition of the four-metric is then given by 
\begin{equation} 
\label{eqn:4met31}
g_{\mu\nu}=h_{\mu\nu}-n_{\mu}n_{\nu},
\end{equation}
where $h_{\mu\nu}$ is the metric induced to the hypersurface
$\Sigma$. The timelike vector $n^{\mu}$ is orthogonal to the
hypersurface $\Sigma$, that is, $h_{\mu\nu}n^{\nu}=0$. This is the
core of the $3+1$ decomposition:
$g^{\mu\alpha}h_{\alpha\nu}=h^{\mu}_{\nu}$ projects the components of
four-tensors to the spatial hypersurface $\Sigma_t$, while $n^{\mu}$
projects them to the direction orthogonal to it. Using
these projections, a four-tensor $T_{\mu\nu}$, for example, can be
decomposed in the following way: 
\begin{align}
\label{eqn:Tprjct}
T_{\mu\nu}&=\left(h^{\alpha}_{\mu}-n^{\alpha}n_{\mu}\right)\left(h^{\beta}_{\nu}-n^{\beta}n_{\nu}\right)T_{\alpha\beta} \nonumber\\
&=\,_{||}\!T_{\mu\nu}-\,_{||}\!T_{\mu\bot}-\,_{||}\!T_{\bot\nu}+T_{\bot\bot},
\end{align}
where ``$||$'' denotes that the greek indices are projected to the
hypersurface using  $h^{\alpha}_{\mu}$, while ``$\bot$'' denotes the
position of an index that has been projected along the orthogonal
vector $n^{\mu}$. 

The decomposition \eqref{eqn:4met31} implies that the four-metric and its determinant decompose as
\begin{equation}
\label{eqn:g31mat}
g_{\mu\nu}=\left(
\begin{matrix}
-N^{2}+N_{i}N^{i}& N_{i}\\[6pt]
N_{i} & h_{ij} &
\end{matrix}
\right)\,,\quad \sqrt{-g}=N\sqrt{h},
\end{equation} 
where $h_{ij}$ is now the three-metric as directly formulated with
spatial indices, which is used to raise and lower spatial indices; we denote $h:={\rm det}\,h_{ij}$. The ``local three-volume'' $\sqrt{h}$ is often referred to as an intrinsic time because it makes the kinetic term of the Hamiltonian indefinite; see, for example \cite{Witt67} and \cite{OUP} [see there in particular Eq.~(5.21)]. The inverse of the four-metric has the form
\begin{equation}
\label{eqn:g31invmat}
g^{\mu\nu}=\left(
\begin{matrix}
-1/N^{2}& N^{i}/N^2\\[6pt]
N^{i}/N^2 & h^{ij}-N^{i}N^{j}/N^2 &
\end{matrix}
\right)\,.
\end{equation}
With these definitions, the time components of objects projected to
the hypersurface vanish; in \eqref{eqn:Tprjct}, for example, all
components with ``$||$'' are now spatial, and the ``$||$'' can be
dropped with the understanding that greek indices can there be turned
into latin ones $i,j$, etc.: $\,_{||}\!T_{\mu\nu}\rightarrow \,^{\ssst
  (3)}\! T_{ij}$, $\,_{||}\!T_{\bot\nu}\rightarrow T_{\bot j}$, etc.,
where objects denoted with a left superscript ``$(3)$'' are intrinsic
to the hypersurface. 

The Riemann tensor, Ricci tensor, and Ricci scalar can be decomposed in a manner similar to \eqref{eqn:Tprjct}. The resulting expressions are well known and can be found, for example, in \cite{OUP,Wald,Pad}; for the decomposition of the Weyl tensor we refer to \cite{Kluson2014} and \cite{ILP} for a derivation. Here, we only state the final expressions for the Ricci scalar and the squared Weyl tensor,
\begin{align}
\label{eqn:Rdec}
&R=\,^{\ssst (3)}R+K_{ij}K^{ij}+K^{2}+2\mathcal{L}_{n}K-\frac{2}{N}D^{i}D_{i}N\,\\
\label{eqn:Rdec1}
&\,\,\,\,=\,^{\ssst (3)}R+K_{ij}K^{ij}-K^{2}+2\nabla_{\mu}\left(n^{\mu}K\right)-\frac{2}{N}D^{i}D_{i}N,\\
\label{eqn:C2dec}
&C_{\mu\nu\lambda\rho}C^{\mu\nu\lambda\rho}=8C_{i\bot j\bot}C^{i\bot j\bot}-4C_{ijk}C^{ijk},
\end{align}
where
\begin{align}
\label{eqn:Kijdef}
K_{ij}&=\frac{1}{2}\mathcal{L}_{n}h_{ij}=\frac{1}{2N}\left(\dot{h}_{ij}-2D_{(i}N_{j)}\right),\\
\label{eqn:Ktrdef}
K&=h^{ij}K_{ij}=\frac{\mathcal{L}_{n}\sqrt{h}}{\sqrt{h}}=\frac{1}{N}\left(\frac{\dot{\sqrt{h}}}{\sqrt{h}}-D_{i}N^{j}\right)
\end{align}
are the extrinsic curvature (second fundamental form), $K=h^{ij}K_{ij}$ its trace, and $D_{i}$ the spatial covariant derivatives with respect to $h_{ij}$. The quantities\footnote{A superscript ``T'' always denotes a traceless object.}
\begin{align}
\label{eqn:elW}
C_{ij}^{\ssst\rm T}&:= -2C_{i\bot j\bot}\nonumber\\
&=\mathbb{1}_{(ij)}^{ab\ssst \rm T}\left(\mathcal{L}_{n}K_{ab}-\!\,^{\ssst (3)}\!R_{ab}-K_{ab}K-\frac{1}{N}D_{ab}N\right)\,,\nonumber\\
&=\left(\mathcal{L}_{n}K_{ij}\right)^{\ssst\rm T}-\!\,^{\ssst
  (3)}\!R_{ij}^{\ssst\rm T}-K_{ij}^{\ssst \rm
  T}K-\frac{1}{N}D_{ij}^{\ssst \rm T}N,\\ 
\label{eqn:magW}
C_{ijk}&:= C_{ijk\bot}=2S_{ijk}^{def}D_{d}K_{ef}\,,
\end{align}
are related to the ``electric'' and ``magnetic'' parts of the Weyl tensor \cite{ILP}, where $\mathbb{1}^{ab\ssst\rm T}_{(ij)}$ and $S_{ijk}^{def}$ are defined as
\begin{align}
\label{eqn:oneT}
\mathbb{1}^{ab\ssst \rm T}_{(ij)}&:= \delta_{(i}^{a}\delta_{j)}^{b}-\frac{1}{3}h_{ij}h^{ab},\\
\label{eqn:SDK}
S_{ijk}^{def}&:=\delta_{i}^{[d}\left(\delta_{j}^{e]}\delta_{k}^{f}-h_{jk}h^{e]f}\right),
\end{align}
and $D_{ij}^{\ssst \rm T}\equiv \mathbb{1}_{(ij)}^{ab\ssst \rm T} D_{ab} \equiv\mathbb{1}_{(ij)}^{ab\ssst \rm T}D_{a}D_{b}$. Note that the term $C_{ij}^{\ssst\rm T}C^{ij\ssst\rm T}$ 
in \eqref{eqn:C2dec} contains only traceless quantities and does not contain velocities of the trace $K$, but contains the trace $K$ itself. This is an important observation to be referred to in our Hamiltonian formulation below.

It is evident from \eqref{eqn:C2dec}, \eqref{eqn:elW}, and \eqref{eqn:Kijdef} that the Lagrangian of the W theory is of second order in the time derivatives of the three-metric and that the order cannot be reduced using partial integration. In order to formulate the theory canonically, one needs to introduce a new variable in order to ``hide'' the first derivative.\footnote{Or the second derivative \cite{Blw,BOS}, but hiding the lower derivatives seems more practical and intuitive.} For this, we add to the original Lagrangian density $\mathcal{L}^{\ssst\rm W}$ a term that implements the relation \eqref{eqn:Kijdef},
\begin{align}
\label{eqn:LagC}
&\mathcal{L}^{\ssst\rm W}\rightarrow\mathcal{L}^{\ssst\rm W}_{c}=\mathcal{L}^{\ssst\rm W}-\lambda^{ij}\left(2K_{ij}-\mathcal{L}_{n}h_{ij}\right)\,,
\end{align}
with Lagrange multipliers $\lambda^{ij}$. This will be the starting point for the Hamiltonian formulation of the Weyl-squared theory.

The addition of the EH term changes nothing regarding the promotion of $K_{ij}$ to a canonical variable, even if the usual boundary term is subtracted from it, as will be discussed in more detail in Sec.~\ref{ClassWE}.

We are now ready for the Hamiltonian formulation of the two theories.

\section{Unimodular-conformal $3+1$ configuration variables}
\label{umod}

The Hamiltonian formulation of Weyl gravity is by no means new. 
The previous works \cite{Kaku1982,Blw,Hwz,Quer,DerrHD,Kluson2014,ILP}
have addressed such a formulation in several ways, some of which are
more similar to each other than others. But there are still some gaps
in understanding the constraint structure and the conformal
symmetry. One of the most important ones is the following. If the trace
of the extrinsic curvature $K$ is indeed an arbitrary object in the
theory (as originally observed by Kaku \cite{Kaku1982} and Boulware
\cite{Blw}), one can  conclude that the local volume element
(intrinsic time) 
$\sqrt{h}$, in which the scale degree of freedom of the three-metric
is contained, should be arbitrary, too, since in the conformally
invariant Weyl-tensor theory all scales are irrelevant. Consequently,
it should be possible to formulate the Hamiltonian constraint in a
conformally \textit{invariant} way. We believe that the formal reason
for not implementing this fact so far lies in not making use of an
irreducible unimodular-conformal decomposition for {\em all} $3+1$
variables. 
It is the purpose of this section to introduce such variables and to use them to
perform the Hamiltonian analysis in the following sections, which will
show that the Hamiltonian constraint can indeed be made conformally
\textit{invariant}. This is also suitable for the discussion of
the quantum theory for which we can observe a connection between
conformal symmetry and the absence of intrinsic time \cite{KN17a,KN17b}. 

In the following, we shall define the unimodular-conformal $3+1$ canonical
variables with which we will separate the full set of canonical
variables into a conformally invariant and a conformally noninvariant
part. The irreducible nature of the unimodular-conformal decomposition
is crucial for this. It was already mentioned in \cite{DerrHD} in
connection with the Hamiltonian formulation of $f({\rm Riem})$
theories that the canonical description of the W theory would be more
transparent if one could isolate the determinant of the three-metric
as a canonical variable. The present paper puts this into practice and
relates the unimodular decomposition of the three-metric to the
conformal decomposition of the extrinsic curvature. Namely, we shall
decompose the three-metric into its scale part, ${(\sqrt{h})}^{\ssst
  2/3}$, and its unimodular (conformal) part, $\bar{h}_{ij}$. The
usefulness of such a unimodular decomposition can be seen in other
situations; see, for example, \cite{Jinn}. We shall, however, go
beyond the decomposition of only the three-metric and decompose also
the lapse function $N$ into its scale part and a scale free lapse
density $\bar{N}$. We observe that the contravariant shift vector
$N^{i}$ is already scale free, that is, conformally invariant, while
its covariant version can be decomposed into a scale part and a
scale free density part $\bar{N}_{i}$. 
In explicit form, the decomposition of lapse, shift, and three-metric reads
\begin{align}
\label{eqn:umoddec}
N^{i}&=:\bar{N}^{i}\,,\quad N_{i}=:{(\sqrt{h})}^{\frac{2}{3}}\bar{N}_{i}\,,\nonumber\\
N&=:(\sqrt{h})^{\frac{1}{3}}\bar{N}\,,\quad h_{ij}=:{(\sqrt{h})}^{\frac{2}{3}} \bar{h}_{ij}\,.
\end{align}
All barred objects, being scale free, are invariant under conformal
transformations (see Appendix \ref{AppConfUm}). This decomposition
then suggests for the hypersurface-orthogonal unit four-vector the
definitions 
\begin{align}
\label{eqn:vecnbar}
n^{\mu}&=: {(\sqrt{h})}^{-\frac{1}{3}}\bar{n}^{\mu}={(\sqrt{h})}^{-\frac{1}{3}}\left(\frac{1}{\bar{N}},-\frac{N^{i}}{\bar{N}}\right),\nonumber\\
n_{\mu}&:= {(\sqrt{h})}^{\frac{1}{3}}\bar{n}_{\mu}={(\sqrt{h})}^{\frac{1}{3}}\left(-\bar{N},0\right),
\end{align}
which further implies
\begin{equation}
\label{eqn:Liebar}
\mathcal{L}_{n}\mathcal{T}={(\sqrt{h})}^{-\frac{1}{3}}\mathcal{L}_{\bar{n}}\mathcal{T}
\end{equation}
for the Lie derivative along $\bar{n}^{\mu}$ with respect to the usual Lie derivative along $n^{\mu}$ of any tensor (density) $\mathcal{T}$.

Using \eqref{eqn:umoddec}, the extrinsic curvature \eqref{eqn:Kijdef} can be decomposed as
\begin{align}
\label{eqn:Kdecomp}
K_{ij}&=\frac{{(\sqrt{h})}^{\frac{1}{3}}}{2\bar{N}}\left(\dot{\bar{h}}_{ij}-2\left[D_{(i}\bar{N}_{j)}\right]^{\ssst\rm T}\right)\nonumber\\
&\quad+\frac{1}{3}h_{ij}\frac{1}{(\sqrt{h})^{\frac{1}{3}}\bar{N}}\left(\frac{\dot{\sqrt{h}}}{\sqrt{h}}-D_{i}\bar{N}^{i}\right)\,,
\end{align}
where the superscript ``$\ssst \rm T$'' denotes that the expression in
the brackets is traceless. Notice from the structure of
\eqref{eqn:Kdecomp} that we can identify explicitly the traceless and
trace parts of $K_{ij}$ (corresponding to ``shear'' and
``expansion''), each of which can be decomposed suitably such 
that the resulting objects have a simplified conformal transformation
law,\footnote{The tracelessness of $\dot{\bar{h}}_{ab}$ can be seen by
  using $\delta h=hh^{ab}\delta h_{ab}$ to show that $h^{ab}\delta \bar{h}_{ab}=0$.}
\begin{align}
\label{eqn:Ksplit}
K_{ij}&=K_{ij}^{\ssst\rm T}+\frac{1}{3}h_{ij}K,\quad {\rm where}\\
\label{eqn:Ktbar}
K_{ij}^{\ssst\rm T}&=\frac{{(\sqrt{h})}^{\frac{1}{3}}}{2\bar{N}}\left(\dot{\bar{h}}_{ij}-2\left[D_{(i}\bar{N}_{j)}\right]^{\ssst\rm T}\right)=:{(\sqrt{h})}^{\frac{1}{3}}\bar{K}_{ij}^{\ssst\rm T},\\
\label{eqn:Ktrbar}
K&=\frac{1}{(\sqrt{h})^{\frac{1}{3}}\bar{N}}\left(\frac{\dot{\sqrt{h}}}{\sqrt{h}}-D_{i}\bar{N}^{i}\right)=: 3{(\sqrt{h})}^{-\frac{1}{3}}\bar{K}.
\end{align}
We thus have arrived at the interesting conclusion that the irreducible unimodular
decomposition of the three-metric induces an irreducible decomposition
of the extrinsic curvature into it traceless and trace parts; to our
knowledge, this has not been remarked before in the
literature. Notice that the trace density $\bar{K}$, unlike the
traceless extrinsic curvature density $\bar{K}_{ij}^{\ssst\rm T}$,
still contains the scale and transforms inhomogeneously under
conformal transformations, see Eq. \eqref{eqn:confK} in Appendix
\ref{AppConfUm}; it represents the evolution of the scale (or local
three-volume). One can also put together \eqref{eqn:Ktbar},
\eqref{eqn:Ktrbar}, and $h_{ij}={(\sqrt{h})}^{\ssst 2/3}\bar{h}_{ij}$
from \eqref{eqn:umoddec} into \eqref{eqn:Ksplit} and write 
\begin{equation}
\label{eqn:Kbarfull}
K_{ij}=: {(\sqrt{h})}^{\frac{1}{3}}\bar{K}_{ij}={(\sqrt{h})}^{\frac{1}{3}}\left(\bar{K}_{ij}^{\ssst\rm T}+\bar{h}_{ij}\bar{K}\right) ,
\end{equation}
where the combination in the parentheses could be referred to as extrinsic curvature density.

The conformal nature of most of the variables resulting from the
unimodular-conformal decomposition with \eqref{eqn:umoddec},
\eqref{eqn:Ktbar}, and \eqref{eqn:Ktrbar} motivates us to choose the
following \textit{canonical} variables for studying
\eqref{eqn:W-action} and its extension \eqref{eqn:WEHm-action}, see
also Appendix \ref{AppConfUm}, 
\begin{align}
\label{eqn:VarsNbar}
\bar{N}^{i}&=N^{i}\,,\\
\bar{N}_{i}&=a^{-2} N_{i}\,,\qquad\bar{N}=a^{-1}N\,,\\
\label{eqn:Varshbar}
\bar{h}_{ij}&=a^{-2}h_{ij}\,,\,\qquad a:={(\sqrt{h})}^{\frac{1}{3}}\,,\\
\label{eqn:VarsKbar}
\bar{K}_{ij}^{\ssst\rm T}&=a^{-1}K_{ij}^{\ssst\rm T}\,,\quad\,\,\, \bar{K}=\frac{aK}{3}\,.
\end{align}
Except $a$ and $\bar{K}$, all the new variables are deprived of scale density
and are thus conformally invariant; $a$ and $\bar{K}$ transform under
conformal transformations as
\begin{align}
\label{eqn:aKbartransf}
a\quad\rightarrow\quad \tilde{a}&=\Omega a\,,\\
\bar{K}\quad\rightarrow\,\,\, \tilde{\bar{K}}&=\bar{K}+\mathcal{L}_{\bar{n}}\log\Omega,
\end{align}
where according to \eqref{eqn:Liebar}
$\mathcal{L}_{\bar{n}}\log\Omega=\bar{n}^{\mu}\del_{\mu}\log\Omega={(\sqrt{h})}^{\ssst -1/3}n^{\mu}\del_{\mu}\log\Omega$, and $\bar{K}_{ij}^{\ssst\rm T}$ and $\bar{K}$ are given by
\begin{align}
\label{eqn:Kbardef}
\bar{K}_{ij}^{\ssst\rm T}&=\frac{1}{2\bar{N}}\left(\dot{\bar{h}}_{ij}-2\left[D_{(i}\bar{N}_{j)}\right]^{\ssst\rm T}\right)\\[6pt]
\label{eqn:Kbardef1}
&=\frac{1}{2\bar{N}}\left(\dot{\bar{h}}_{ij}-2\left[\bar{D}_{(i}\bar{N}_{j)}\right]^{\ssst\rm T}\right),\\[6pt]
\label{eqn:Ktbardef}
\bar{K}&=\frac{1}{\bar{N}}\left(\frac{\dot{a}}{a}-\frac{1}{3}D_{i}N^{i}\right)\\[6pt]
\label{eqn:Ktbardef1}
&=\frac{\bar{n}^{\mu}\del_{\mu}a}{a}-\frac{\del_{i}N^{i}}{3\bar{N}}.
\end{align}
Equations \eqref{eqn:Kbardef1} and \eqref{eqn:Ktbardef1} were derived using the results from 
Appendix~\ref{AppRicciBar}, and they manifestly reveal that the extrinsic curvature transforms inhomogeneously due to the first term in \eqref{eqn:Ktbardef1}, while \eqref{eqn:Kbardef1} is conformally invariant. Note that the ``bar derivative'' $\bar{D}_i$ is defined with respect to the conformal part of the metric $\bar{h}_{ij}$ and is not of covariant nature (see Appendix~\ref{AppRicciBar}). It should be noted that all new variables are now tensor densities, except the contravariant shift vector $\bar{N}^{i}=N^{i}$, whose bar we omit from now on.

Note that similar decompositions are used in a number of instances related, in particular, to the Cauchy problem \cite{SN,BS} and to conformal-traceless decompositions of Einstein equations \cite{DB}. We refer to $\bar{N}$, $\bar{K}_{ij}^{\ssst\rm T}$, and $\bar{K}$ as the ``lapse density'',\footnote{Also called ``densitized lapse'' or ``Taub function'' \cite{RJ}.} ``traceless extrinsic curvature density'', and ``trace density'', respectively.

One now expects that any conformally invariant theory --- such as the
W theory --- will be deprived of scale density $a$ and trace density
$\bar{K}$. Moreover, any conformal symmetry breaking of a theory would
be connected to only these two variables. In other words, their
absence reflects conformal invariance. Accordingly, we expect
the configuration space structure to be much simpler and the
constraints easier to understand with the choice of
unimodular-conformal variables as \textit{canonical} variables. This
is investigated in the following sections.

\section{Pure Weyl-squared gravity}
\label{ham}

\subsection{Hamiltonian formulation and constraint analysis}
\label{ClassW}

The starting point is the Weyl-tensor action \eqref{eqn:W-action}, for which the $3+1$ decomposition \eqref{eqn:C2dec} of the Weyl tensor as well as the added constraint in \eqref{eqn:LagC} are used. 
 Recall that the latter is needed to take care of the additional degrees of freedom introduced by promoting $K_{ij}$ to an independent variable. This leads to the action
\begin{align}
\label{eqn:W31}
S^{\ssst \rm W}=\inttx\,N\sqrt{h}\Biggl\lbrace&-\frac{\alpha_{\ssst\rm W}}{2} C_{ij}^{\ssst\rm T}C^{ij\ssst\rm T}+\alpha_{\ssst\rm W} C_{ijk}C^{ijk}\nonumber\\
&-\lambda^{ij}\left(2K_{ij}-\mathcal{L}_{n}h_{ij}\right)\Biggr\rbrace.
\end{align}
Unlike previous works, we now implement the
unimodular-conformal variables introduced in the last section. Term by
term, this leads us to the following expressions. 

Using in \eqref{eqn:elW} Eq.~\eqref{eqn:tLieKtelW} from Appendix~3, the electric part of the Weyl tensor becomes
\begin{equation}
\label{eqn:elWsmpl}
C_{ij}^{\ssst\rm T}=\mathcal{L}_{n}K_{ij}^{\ssst\rm T}-\frac{1}{3}K_{ij}^{\ssst \rm T}K-\frac{2}{3}h_{ij}K_{ab}^{\ssst \rm T}K^{ab\ssst \rm T}-\!\,^{\ssst (3)}\!R_{ij}^{\ssst\rm T}-\frac{1}{N}D_{ij}^{\ssst \rm T}N,
\end{equation}
where $D_{j}a=0$ was used. Using \eqref{eqn:Liebar}, the rightmost
identity in \eqref{eqn:Ktbar}, and \eqref{eqn:Ktrdef}, the first two
terms on the right-hand side of \eqref{eqn:elWsmpl} reduce to 
\begin{equation}
\label{eqn:elWLiered}
\mathcal{L}_{n}K_{ij}^{\ssst\rm T}-\frac{1}{3}K_{ij}^{\ssst \rm T}K=\mathcal{L}_{\bar{n}}\bar{K}_{ij}^{\ssst\rm T}.
\end{equation}
Thus not only the velocity $\dot{K}$, but also the trace of the
extrinsic curvature itself disappears explicitly from the Lagrangian
of the Weyl-squared theory.\footnote{This holds up to a possible
  boundary term: we expect that $K$ either vanishes from
  $C_{ijk}C^{ijk}$ or occurs only in a total divergence, but we do not
  attempt to prove this here.} Furthermore, the scale density $a$ disappears
as well, because $\,^{\ssst (3)}\! R_{ij}^{\ssst\rm
  T}+\frac{1}{N}D_{ij}^{\ssst \rm T}N$ can be shown not to depend on
$a$ (see Appendix~\ref{AppIdent}). In fact, $K$ and $a$ were never
there in the first place, owing to the conformally invariant nature of
the Weyl tensor, but this fact is obscured if   the original variables
are used. This is directly related to the fact that the Weyl tensor is
traceless. One can thus make the following statement: \textit{In the
  Weyl-squared theory, traces and scales do not propagate and should
  thus not appear explicitly in the constraints resulting from the
  Hamiltonian formulation} -- they are arbitrary.  

We will write an overbar to $C_{ij}^{\ssst\rm T}$, $C_{ij}^{\ssst\rm
  T}\equiv\bar{C}_{ij}^{\ssst\rm T}$ in order to mark that it is
expressed in terms of the new variables, 
\begin{equation}
\label{eqn:elWumod}
\bar{C}_{ij}^{\ssst\rm T}=\mathcal{L}_{\bar{n}}\bar{K}_{ij}^{\ssst\rm
  T}-\frac{2}{3}\bar{h}_{ij}\bar{K}_{ab}^{\ssst \rm
  T}\bar{h}^{an}\bar{h}^{bm}\bar{K}_{nm}^{\ssst \rm T}-\!\,^{\ssst
  (3)}\!\bar{R}_{ij}^{\ssst\rm
  T}-\frac{1}{\bar{N}}\left[\bar{D}_{i}\del_{j}\bar{N}\right]^{\ssst\rm
  T}\,. 
\end{equation}
Note that the trace of the first term on the right-hand side of
\eqref{eqn:elWumod} is canceled by the second term, leaving only the
traceless part of $\mathcal{L}_{\bar{n}}\bar{K}_{ij}^{\ssst\rm T}$,
while the combination $\bar{R}_{ij}^{\ssst\rm
  T}+\frac{1}{\bar{N}}\left[\bar{D}_{i}\del_{j}\bar{N}\right]^{\ssst\rm
  T}$ is conformally invariant; see
\eqref{eqn:DDphi}--\eqref{eqn:DDRTeq}. 

Using \eqref{eqn:Kbarfull} and \eqref{eqn:magW},
the $C_{ijk}^2$-term in \eqref{eqn:W31} is seen to scale with $a^{\ssst -4}$, 
\begin{align}
\label{eqn:magWumod}
C_{ijk}C^{ijk}&=C_{ijk}h^{ia}h^{jb}h^{kc}C_{abc}\nonumber\\
&=a^{-4}\bar{C}_{ijk}\bar{h}^{ia}\bar{h}^{jb}\bar{h}^{kc}\bar{C}_{abc}\nonumber\\
&\equiv a^{-4}\bar{C}_{ijk}^2\,,\quad {\rm for\,\,short}.
\end{align}

In the last term of the Lagrangian in \eqref{eqn:W31}, we simply split all objects, using \eqref{eqn:Kdecomp} and \eqref{eqn:VarsNbar}--\eqref{eqn:VarsKbar}, as well as \eqref{eqn:Kbardef1} and \eqref{eqn:Ktbardef1}, and finally obtain the following Lagrangian, which is the starting point for our canonical formalism,
\begin{align}
\label{eqn:LcW}
\mathcal{L}_{c}^{\ssst\rm W}&=\bar{N}\Biggl\lbrace-\frac{\alpha_{\ssst\rm W}}{2} \bar{h}^{ia}\bar{h}^{jb}\bar{C}_{ij}^{\ssst\rm T}\bar{C}_{ab}^{\ssst\rm T}+\alpha_{\ssst\rm W} \bar{C}_{ijk}^{2}\nonumber\\
&\qquad\,\,\,\,-a^{5}\lambda^{ij\ssst\rm T}\left[2\bar{K}_{ij}^{\ssst\rm T}-\frac{1}{\bar{N}}\left(\dot{\bar{h}}_{ij}-2\left[\bar{D}_{(i}\bar{N}_{j)}\right]^{\ssst\rm T}\right)\right]\nonumber\\
&\qquad\,\,\,\,-2a^3\lambda \left[\bar{K}-\frac{1}{\bar{N}}\left(\frac{\dot{a}}{a}-\frac{1}{3}D_{a}N^{a}\right)\right]\Biggr\rbrace\,.
\end{align}
We note that the scale density $a={(\sqrt{h})}^{\ssst 1/3}$ and the trace $\bar{K}$ have vanished from the  Weyl-tensor part of this Lagrangian, as is expected for a conformally invariant theory. It then seems unnecessary to introduce $\bar{K}$ as an independent variable, but since we want to start from the full configuration space, not the subspace, we will take the full $K_{ij}$ as independent degrees of freedom. This provides a deep insight into the structure of the theory.

The canonical momenta conjugate to our unimodular-conformal variables are then defined as follows:
\begin{align}
\label{eqn:pNbar}
p_{\ssst \bar{N}}&=\frac{\del \mathcal{L}^{\ssst\rm W}_{c}}{\del \dot{\bar{N}}}\approx 0\,,\quad\quad p_{i}=\frac{\del \mathcal{L}^{\ssst\rm W}_{c}}{\del \dot{\bar{N}}^{i}}\approx 0\,,\\[8pt]
\label{eqn:phbar}
\bar{p}^{ij}&=\frac{\del \mathcal{L}^{\ssst\rm W}_{c}}{\del \dot{\bar{h}}_{ij}}=a^{5}\lambda^{ij \ssst\rm T}\,,\\[8pt]
\label{eqn:pa}
p_{a}&=\frac{\del \mathcal{L}^{\ssst\rm W}_{c}}{\del \dot{a}}=2a^2 \lambda\,,\\[8pt]
\label{eqn:pKtbar}
\bar{P}^{ij}&=\frac{\del \mathcal{L}^{\ssst\rm W}_{c}}{\del \dot{\bar{K}}_{ij}^{\ssst\rm T}}=-\alpha_{\ssst\rm W}\,\bar{h}^{ia}\bar{h}^{jb}\bar{C}_{ab}^{\ssst\rm T}\,,\\[6pt]
\label{eqn:pKbar}
\bar{P} &=\frac{\del \mathcal{L}^{\ssst\rm W}_{c}}{\del \dot{\bar{K}}}\approx 0\,,
\end{align}
where the ``$\approx$'' is Dirac's ``weak equality'' \cite{Dir}. It
can be shown \cite{Kluson2014} that it is unnecessary to include
variables $\lambda_{ij}^{\ssst\rm T},\lambda $ and their conjugate
momenta as canonical variables, and then \eqref{eqn:phbar} and
\eqref{eqn:pa} are strong equalities. Note that the momenta
\eqref{eqn:phbar}---\eqref{eqn:pKtbar} are tensor
densities of scale weight\footnote{The ``scale weight'' $w_a$, which
  is related to the weight $w$ of a tensor density by $w_{a}=3 w$, is
  introduced in Appendix~\ref{AppRicciBar}. } $w_{a}=5$, $w_{a}=2$,
and $w_{a}=4$, respectively. Note that the momenta \eqref{eqn:phbar}
and \eqref{eqn:pKtbar} are traceless. One can conclude from the above
that there are three primary constraints of which 
\begin{equation}
\label{eqn:Pcnstr}
\bar{P}\approx 0
\end{equation}
is a new one compared to the standard ADM formulation of GR where we
only have the two corresponding to \eqref{eqn:pNbar} \cite{OUP}. This
new constraint is signaling the arbitrariness of $\bar{K}$ (and
implicitly $K$) if it turns out (as it will) to be a first class
constraint. Within the formalism of \cite{Kluson2014}, where the
original variables are used, one cannot conclude that the trace $K$ is
arbitrary because $P$ and $K$ are there \textit{not} canonical
variables, but a combination of them, namely $P=h_{ij}P^{ij}$ and
$K=h^{ij}K_{ij}$, so they cannot be identified one to one with an
arbitrary degree of freedom. Here, however, $\bar{P}\approx 0$ is
similar in nature to \eqref{eqn:pNbar}: if $p_{\bar{N}}$ and $p^{i}$
imply that the lapse $\bar{N}$ and shift $\bar{N}^i$ do not appear in
the constraints, then $\bar{P}\approx 0$ implies that $\bar{K}$ should
not appear, too. Hence, $\bar{K}$ is an arbitrary degree of
freedom. Recalling the definition \eqref{eqn:Ktrbar} of $\bar{K}$, we
expect that $\dot{a}$ should not appear in the constraints as
well. This means that the scale density $a$ should have a vanishing
momentum, too,\footnote{In terms of the original variables, this would
  correspond to $p=h_{ij}p^{ij}\approx 0$, which obviously did not
  appear as one of our primary constraints. In earlier works,
  e.g. \cite{Kluson2014, ILP}, one can find such a term only as a part
  of their conformal (or dilatational) secondary constraint.} 
that is, $p_{a}\approx 0$. But where is that constraint if it does not
appear in the set of primary constraints
\eqref{eqn:pNbar}--\eqref{eqn:pKtbar}? We derive the Hamiltonian
before answering this question. 

Based on the transformation from the original to unimodular-conformal variables, the Poisson brackets with respect to the new variables read
\begin{align}
\label{eqn:PB}
&\left\lbrace A({\mathbf x}), B({\mathbf y})\right\rbrace \nonumber\\
&\quad=\int{\rm d}^3z\Biggl(\frac{\delta A({\mathbf x})}{\delta
  \bar{h}_{ij}({\mathbf z})}\frac{\delta B({\mathbf y})}{\delta
  \bar{p}^{ij}({\mathbf z})}\nonumber\\[6pt]
&-\frac{\delta A}{\del 
  \bar{p}^{ij}}\frac{\delta B}{\delta \bar{h}_{ij}}+\frac{\delta A}{\delta
  a}\frac{\delta B}{\delta p_{a}}-\frac{\delta A}{\delta p_{a}}\frac{\delta
  B}{\delta a}\nonumber\\[6pt] 
&\quad+\frac{\delta A}{\delta \bar{K}_{ij}^{\ssst\rm T}}\frac{\delta B}{\delta 
  \bar{P}^{ij}}-\frac{\delta A}{\delta \bar{P}^{ij}}\frac{\delta B}{\delta
  \bar{K}_{ij}^{\ssst\rm T}}+\frac{\delta A}{\delta \bar{K}}\frac{\delta
  B}{\delta \bar{P}}-\frac{\delta A}{\delta \bar{P}}\frac{\delta B}{\delta
  \bar{K}}\nonumber\\[6pt] 
&\quad+\frac{\delta A}{\delta \bar{N}}\frac{\delta B}{\delta
  p_{\ssst\bar{N}}}-\frac{\delta A}{\delta p_{\ssst\bar{N}}}\frac{\delta
  B}{\delta \bar{N}}+\frac{\delta A}{\delta N^{i}}\frac{\delta B}{\delta
  p_i}-\frac{\delta A}{\delta p_i}\frac{\delta B}{\delta N^i}\Biggr), 
\end{align}
where we have explicitly spelled out the dependence on the spatial
coordinates only for the first term under the integral. 

The transformation \eqref{eqn:VarsNbar}--\eqref{eqn:VarsKbar} is, in fact,
a \textit{canonical} one. The Poisson brackets among the canonical
variables are given by 
\begin{align}
\label{eqn:PBvars1}
\left\lbrace q_{ij}^{A}({\mathbf x}),\Pi^{ab}_{B}({\mathbf y})
  \right\rbrace=\mathbb{1}^{ab\ssst\rm T}_{ij}\delta_{B}^{A}\delta
  ({\mathbf x},{\mathbf y}) 
\end{align}
for the conformally invariant pairs $q_{ij}^{A}=(
\bar{h}_{ij},\bar{K}_{ij}^{\ssst\rm T}),\,\Pi^{ab}_{B}=(
\bar{p}^{ab},\bar{P}^{ab})$, and 
\begin{align}
\label{eqn:PBvars2}
\left\lbrace q^{A}({\mathbf x}),\Pi_{B}({\mathbf y})
  \right\rbrace=\delta_{B}^{A}\delta ({\mathbf x},{\mathbf y}) 
\end{align}
for the scale and trace pairs $q^{A}=(a, \bar{K})\,,\Pi_{B}=(p_{a},
\bar{P})$. Recall that $\mathbb{1}^{ab\ssst\rm T}_{ij}$ is defined in
\eqref{eqn:oneT}; its presence guarantees that both sides of
\eqref{eqn:PBvars1} are traceless, that is, compatible with each other.  
Expressions similar to \eqref{eqn:PBvars1} and \eqref{eqn:PBvars2}
hold for lapse, shift, and their canonical momenta, while
all other Poisson brackets vanish. 

In terms of the canonical variables, the constrained Lagrangian
\eqref{eqn:LcW} reads 
\begin{align}
\label{eqn:WLagCan}
\mathcal{L}_{c}^{\ssst\rm W}=& \bar{N}\biggl[-\frac{\bar{h}_{ia}\bar{h}_{jb}\bar{P}^{ij}\bar{P}^{ab}}{2\alpha_{\ssst\rm W}} -2\bar{K}_{ij}^{\ssst\rm T}\bar{p}^{ij}-a\bar{K}p_{a}+\alpha_{\ssst\rm W}\bar{C}_{ijk}^2\biggr]\nonumber\\
&+\dot{\bar{h}}_{ij}\bar{p}^{ij}+\dot{a}\,p_{a}-2 \bar{D}_{i}\bar{N}_{j}\bar{p}^{ij}+\frac{1}{3}a\,p_{a}D_{i}N^{i}\,,
\end{align} 
where the symmetrization and the ``${\ssst \rm T}$'' were dropped in
the next-to-last term because $\bar{p}^{ij}$ is symmetric and
traceless. The fact that $\alpha_{\ssst\rm W}$ does enter the various
terms in a different way, has important consequences for the
quantum theory \cite{KN17a,KN17b}.

For the Hamiltonian, we need in addition to express
$\dot{\bar{K}}_{ij}^{\ssst\rm T}\bar{P}^{ij}$ in terms of the new
canonical pairs. Using \eqref{eqn:elWumod} and
\eqref{eqn:pKtbar},\footnote{Note that the second term on the
  right-hand side of \eqref{eqn:elWumod} vanishes when contracted with
  the traceless $\bar{P}^{ij}$.} we get 
\begin{align}
\label{eqn:KtPt}
\dot{\bar{K}}_{ij}^{\ssst\rm T}\bar{P}^{ij}&=\bar{N}\biggl[-\frac{\bar{h}_{ia}\bar{h}_{jb}\bar{P}^{ij}\bar{P}^{ab}}{\alpha_{\ssst\rm W}}+\left(\,^{\ssst (3)}R_{ij}+D_{i}D_{j}\right)^{\ssst\rm T}\bar{P}^{ij}\nonumber\\
&\qquad\quad+\mathcal{L}_{\vec{N}}\bar{K}_{ij}^{\ssst\rm
  T}\biggr]+D_{i}\left(D_{j}\bar{N}\bar{P}^{ij}-\bar{N}D_{j}\bar{P}^{ij}\right), 
\end{align}
where $\mathcal{L}_{\vec{N}}$ is the Lie derivative with respect to
the shift vector $N^{i}$, and we have used the Leibniz rule twice to
avoid the double covariant derivative for the lapse. From the
traceless nature of $\bar{P}^{ij}$ it is clear that only the traceless
part of the parentheses contributes, which is denoted
by attaching the superscript ``$\ssst {\rm T}$''.  

The total Hamiltonian is found by a Legendre transform of the constrained Lagrangian supplemented by all primary constraints,
\begin{align}
\label{eqn:WHam}
H^{\ssst\rm W}=&\int{\rm d}^3 x\Biggl\lbrace \dot{\bar{h}}_{ij}\bar{p}^{ij}+\dot{a}p_{a}+\dot{\bar{K}}_{ij}^{\ssst\rm T}\bar{P}^{ij}-\mathcal{L}_{c}^{\ssst\rm W}\nonumber\\
&\qquad\qquad\qquad\qquad\qquad+\lambda_{\ssst\bar{N}}p_{\ssst\bar{N}}+\lambda_{i}p^{i}+\lambda_{\ssst\bar{P}}\bar{P}\Biggr\rbrace\nonumber\\
=&\int{\rm d}^3 x\Biggl\lbrace \bar{N}\Biggl[-\frac{\bar{h}_{ik}\bar{h}_{jl}\bar{P}^{ij}\bar{P}^{kl}}{2\alpha_{\ssst\rm W}}+\left(\,^{\ssst (3)}\! \bar{R}_{ij}^{\ssst\rm T}+\del_{i}\bar{D}_{j}\right)\bar{P}^{ij}\nonumber\\
&\qquad\qquad\quad+2\bar{K}_{ij}^{\ssst\rm T}\bar{p}^{ij}+a\bar{K}p_{a}-\alpha_{\ssst\rm W}\bar{C}_{ijk}^2\Biggr]\nonumber\\
&\qquad\quad+N^{i}\Biggl[-2\bar{D}_{k}
\left(\bar{h}_{ij}\bar{p}^{kj}\right)-\frac{1}{3}D_{i}\left(a\,p_{a}\right)\nonumber\\
&\qquad\qquad\qquad-2\bar{D}_{k}\left(\bar{K}_{ij}^{\ssst\rm T}\bar{P}^{jk}\right)+\bar{P}^{jk}\bar{D}_{i}\bar{K}_{jk}^{\ssst\rm T}\Biggr]\nonumber\\
&\qquad\quad+\lambda_{\ssst\bar{N}}p_{\ssst\bar{N}}+\lambda_{i}p^{i}+\lambda_{\ssst\bar{P}}\bar{P}\Biggr\rbrace+H_{\rm surf},
\end{align}
where $H_{\rm surf}$ are the surface terms arising from partial integration,
\begin{align}
\label{eqn:HWsurf}
H_{\rm surf}&=2\int{\rm d}^3x\left(\del_{i}\left(\bar{N}_{j}\bar{p}^{ij}\right)+\del_{j}\left(\bar{K}_{ia}^{\ssst\rm T}\bar{P}^{ij}N^{a}\right)\right)\nonumber\\
&\quad+\int{\rm d}^3x\,\del_{i}\left(\del_{j}\bar{N}\bar{P}^{ij}-\bar{N}\bar{D}_{j}\bar{P}^{ij}\right)\,.
\end{align}
Except for the term $D_{i}\left(a\,p_{a}\right)$, all covariant derivatives
 in \eqref{eqn:WHam} reduce to barred derivatives $\bar{D}_{i}$, and
 $(\,^{\ssst (3)}\! R_{ij}+D_{i}D_{j})^{\ssst\rm T}\bar{P}^{ij}$
 reduces to $(\,^{\ssst (3)}\! \bar{R}_{ij}^{\ssst\rm
   T}+\del_{i}\bar{D}_{j})\bar{P}^{ij}$ when using
 \eqref{eqn:Gammabar}---\eqref{eqn:Clogaprop} from Appendix
 \ref{AppRicciBar}. This concludes our preparation for the constraint
 analysis.   

Let us now address the constraints. At first glance, one could
interpret the expressions in front of $\bar{N}$ and $N^{i}$ in
\eqref{eqn:WHam} as the Hamiltonian and momentum constraints,
respectively, arising from the conservation of the primary constraints
\eqref{eqn:pNbar} in time. But let us be more careful. We note that
the scale density $a$ appears only in two terms in \eqref{eqn:WHam}. One of
these terms (the term $a\bar{K}p_{a}$) is the only term containing the trace
$\bar{K}$. At this point, it seems to be part of the Hamiltonian
constraint. But since $\bar{P}\approx 0$ implies that $\bar{K}$ is
arbitrary and should vanish from the constraints explicitly, as stated
above, $\bar{K}$ should be (part of) a Lagrange
multiplier.\footnote{At this point, Kaku's prescription
  \cite{Kaku1982} would be to isolate terms with $\bar{K}$ and
  interpret it as a \textit{a Lagrange multiplier times a constraint}
  where the constraint stems from $\dot{\bar{P}}$. Having achieved
  this, Kaku employed his argument that $K$ is arbitrary and arrived
  at his ``dilatational constraint''. A similar treatment can be found
  in \cite{DerrHD}.} Here, this fact follows from the conservation of
the constraint \eqref{eqn:pKbar} in time, 
\begin{equation}
\label{eqn:Pdot}
\dot{\bar{P}}=\left\lbrace\bar{P},H^{\ssst\rm W}\right\rbrace=-\frac{\delta H^{\ssst\rm W}}{\delta \bar{K}}=-\bar{N}a\, p_{a}\stackrel{!}{\approx} 0,
\end{equation}
which introduces a secondary constraint $\mathcal{Q}^{\ssst\rm W}$ given by
\begin{equation}
\label{eqn:QW}
\mathcal{Q}^{\ssst\rm W}:=a p_{a}\approx 0.
\end{equation}
One could, of course, have concluded from this that we have
$p_{a}\approx 0$ instead of $a p_{a}\approx 0$, since we know nothing
about $a$ as a configuration variable, while $p_{a}$ is somewhat
artificial. This would indicate that $a$ is arbitrary, as suspected,
but since the expression is \textit{weakly} vanishing, we keep it as a
whole. Another reason to support taking $ap_{a}$ as the constraint is
that it will turn out to be part of the generator of conformal
transformations; see below. We call the constraint \eqref{eqn:QW}
the ``scaling constraint'', because it implies the conformal
transformation of the scale only, as we shall see, and differs from
the corresponding constraint derived in \cite{Kaku1982, Blw, DerrHD,
  Kluson2014, ILP} and called there the ``conformal (or dilatational)
constraint''. The conformal constraint in these earlier papers
contains an additional term of the form $K_{ij}P^{ij}$, which in our
case is absent because the use of conformally invariant variables
leaves in \eqref{eqn:QW} only those variables that are affected by a
conformal transformation. 

Note that the consistency condition \eqref{eqn:Pdot} is another way of
stating that the total Hamiltonian should not depend on
$\bar{K}$. This is very important because this statement is here
formally realized by the use of the unimodular-conformal canonical
variables; in the original variables used in most of the earlier
work, this conclusion cannot be drawn. 

The demand for the temporal conservation of $\mathcal{Q}^{\ssst\rm W}$
gives no further constraints,  
\begin{equation}
\label{eqn:QWdot}
\dot{Q}^{\ssst\rm W}=\dot{(a p_{a})}=\left\lbrace ap_{a},H^{\ssst\rm W}\right\rbrace = 0.
\end{equation}
In addition, the Poisson bracket between $\mathcal{Q}^{\ssst\rm W}$
and $\bar{P}$ trivially vanishes because $a$, $p_{a}$ and $\bar{P}$
are independent variables. The result \eqref{eqn:QWdot} may also be
understood from the fact that 
\begin{align}
\dot{a}&=\left\lbrace a,H^{\ssst\rm W}\right\rbrace = a\left(\bar{N}\bar{K}+ \frac{1}{3}D_{i}N^{i}\right)\nonumber\\
&=-a\frac{\dot{p}_{a}}{p_{a}}\quad\Rightarrow\quad a\,p_{a}=f({\mathbf x}),
\end{align}
meaning that $ap_{a}$ is a constant of motion. Note that this
calculation is done completely at the level of Poisson brackets,
without imposing any constraints. This result is valid only in the
pure Weyl theory (see the next subsection).  

By examining the two terms in \eqref{eqn:WHam} that contain $ap_a$,
\eqref{eqn:QW} suggests that $\bar{N}\bar{K}+D_{i}N^{i}/3$ is the
Lagrange multiplier (using one of the surface terms and partial
integration on the second of the two terms) for $\mathcal{Q}^{\ssst\rm W}$, so we
can isolate
$\left(\bar{N}\bar{K}+D_{i}N^{i}/3\right)\mathcal{Q}^{\ssst\rm W}$
from the rest of the terms. 
Recalling \eqref{eqn:Ktrbar}, one can easily see that this Lagrange
multiplier is effectively $\dot{a}/a$. One may then conclude that we
do not need to add the secondary constraint $Q^{\ssst \rm W}$ to the
total Hamiltonian by hand with an additional Lagrange multiplier, as
was done in \cite{Kluson2014}; we merely need to isolate it from the
rest of the Hamiltonian, with the Lagrange multiplier essentially
being $\bar{K}$. Such a prescription was used in
\cite{Kaku1982}. Moreover, adding it by hand would break the
equivalence with the Lagrange formulation \cite{Pons88,PSS1997}. 

Demanding that the primary constraints \eqref{eqn:pNbar} be preserved
in time, and keeping in mind that $a p_{a}\approx 0$, we find that the
Hamiltonian and the momentum constraints are given by (adding
$\mathcal{Q}^{\ssst\rm W}$ for completeness) 
\begin{align}
\label{eqn:HamWcf}
\mathcal{H}^{\ssst\rm W}_{\bot}&=-\frac{\bar{h}_{ik}\bar{h}_{jl}\bar{P}^{ij}\bar{P}^{kl}}{2\alpha_{\ssst\rm W}}+\left(\,^{\ssst (3)}\! \bar{R}_{ij}^{\ssst\rm T}+\del_{i}\bar{D}_{j}\right)\bar{P}^{ij}\nonumber\\[6pt]
&\quad+2\bar{K}_{ij}^{\ssst\rm T}\bar{p}^{ij}-\alpha_{\ssst\rm W}\bar{C}_{ijk}^2\approx 0,\\[12pt]
\label{eqn:MomWcf}
\mathcal{H}^{\ssst\rm W}_{i}&=-2\bar{D}_{k}\left(
\bar{h}_{ij}\bar{p}^{jk}\right)-2\bar{D}_{k}\left(\bar{K}_{ij}^{\ssst\rm T}\bar{P}^{jk}\right)+\bar{P}^{jk}\bar{D}_{i}\bar{K}_{jk}^{\ssst\rm T}\\
\label{eqn:MomWcf1}
&=-2\del_{k}\left(
\bar{h}_{ij}\bar{p}^{jk}\right)+\del_{i}\bar{h}_{jk}\bar{p}^{jk}\nonumber\\[6pt]
&\quad-2\del_{k}\left(\bar{K}_{ij}^{\ssst\rm T}\bar{P}^{jk}\right)+\del_{i}\bar{K}_{jk}^{\ssst\rm T}\bar{P}^{jk}\approx 0,\\[12pt]
\label{eqn:Qw}
\mathcal{Q}^{\ssst\rm W}&=ap_{a}\approx 0.
\end{align}
Here, we have used \eqref{eqn:Gammabar} and \eqref{eqn:Cloga} and the density nature of the variables to reduce the momentum constraint from \eqref{eqn:MomWcf} to \eqref{eqn:MomWcf1}.
It can now easily be seen that no scale density $a$ or trace density $\bar{K}$ appear in the Hamiltonian and momentum constraints. Every object in these two constraints is conformally invariant, which confirms our earlier claim that no constraint should depend on the trace $\bar{K}$. The conformal constraint has to contain the scale density because it carries information about the conformal transformation, as we discuss below.

Thus one can conclude that the ``intrinsic time'' $a$ is absent from
the W theory, along with its time evolution encoded in $\bar{K}$. In
the case of GR, the Hamiltonian constraint is much simpler, and one
cannot get rid of the scale density $a$. The momentum constraints in the
W theory have a structure similar to the one in GR, and additional
terms are due to the phase space being extended by the canonical pair
$(\bar{K}_{ij}^{\ssst\rm T},\bar{P}^{ij})$. The most important
difference is that while the extrinsic curvature leads in GR to the
momentum conjugate to the three-metric, it is in the W theory an
independent canonical variable. 

\subsection{Constraint algebra and number of degrees of freedom}

Using unimodular-conformal variables, the algebra of constraints is
expected to be simple. We do not give a proof for the first three
Poisson brackets below, but take the results from \cite{DerrHD}, where
it is shown that the hypersurface algebra for a \textit{general}
$f({\rm Riem})$ theory is the same as for GR. This can be understood
as a consequence of the reparametrization invariance for such theories
(\cite{Thiemann}, Sec.~1.5). Writing the smeared versions of the
constraints \eqref{eqn:HamWcf}--\eqref{eqn:MomWcf1} and
\eqref{eqn:Pcnstr} 
(summarized as $\mathcal{C}^{A}$) as
\begin{equation}
\mathcal{C}^{A}[\eta]=\intx \,\eta ({\mathbf x})\cdot
\mathcal{C}^{A}({\mathbf x}), 
\end{equation}
where $\eta ({\mathbf x})$ is an arbitrary (vector or scalar)
function, the Poisson brackets among them are 
\begin{align}
\label{eqn:AlgHH}
\left\lbrace\mathcal{H}_{\bot}^{\ssst\rm W}[\varepsilon_1],\mathcal{H}_{\bot}^{\ssst\rm W}[\varepsilon_2]\right\rbrace &=\mathcal{H}_{||}^{\ssst\rm W}[\varepsilon_1 \del^{i}\varepsilon_2-\varepsilon_2 \del^{i}\varepsilon_1]\,,\\[2pt]
\label{eqn:AlgHm}
\left\lbrace\mathcal{H}_{||}^{\ssst\rm W}[\vec{\eta}],\mathcal{H}_{\bot}^{\ssst\rm W}[\varepsilon]\right\rbrace &=\mathcal{H}_{\bot}^{\ssst\rm W}[\mathcal{L}_{\vec{\eta}}\varepsilon]\,,\\[2pt]
\label{eqn:Algmm}
\left\lbrace\mathcal{H}_{||}^{\ssst\rm W}[\vec{\eta}_1],\mathcal{H}_{||}^{\ssst\rm W}[\vec{\eta}_2]\right\rbrace &=\mathcal{H}_{||}^{\ssst\rm W}[\mathcal{L}_{\vec{\eta}_{1}}\vec{\eta}_2]\,,\\[2pt]
\label{eqn:AlgHP}
\left\lbrace\mathcal{H}_{\bot}^{\ssst\rm W}[\varepsilon],\bar{P}[\epsilon]\right\rbrace &=0\,,\\[2pt]
\label{eqn:AlgmP}
\left\lbrace\mathcal{H}_{||}^{\ssst\rm W}[\vec{\eta}],\bar{P}[\epsilon]\right\rbrace &=0\,,\\[2pt]
\label{eqn:AlgHQ}
\left\lbrace\mathcal{H}_{\bot}^{\ssst\rm W}[\varepsilon],\mathcal{Q}^{\ssst\rm W}[\omega]\right\rbrace &=0\,,\\[2pt]
\label{eqn:AlgHmQ}
\left\lbrace\mathcal{H}_{||}^{\ssst\rm W}[\vec{\eta}],\mathcal{Q}^{\ssst\rm W}[\omega]\right\rbrace &=0\,,\\[2pt]
\label{eqn:AlgPQ}
\left\lbrace \bar{P}[\epsilon],\mathcal{Q}^{\ssst\rm W}[\omega]\right\rbrace &=0,
\end{align}
and all constraints are first class. We note, however, that in
\cite{Kluson2014} and \cite{ILP} the foliation algebra
\eqref{eqn:AlgHH}--\eqref{eqn:Algmm} contains an additional term of
the form $P[(\varepsilon_1 D^{i}\varepsilon_2-\varepsilon_2
D^{i}\varepsilon_1)(D_{j}{K^{j}}_{i}-D_{i}K)]$. The consequence of its
presence is unclear, but its origin lies in the fact that the
$P$-constraint was there not taken into account when defining the
Hamiltonian, momentum, and conformal constraints. Using
unimodular-conformal variables, in which neither $\bar{P}$ nor
$\bar{K}$ enter any of the secondary constraints, it is not surprising
that our results for the foliation algebra give those of
\cite{DerrHD}. Since the transformation to the unimodular-conformal
variables is canonical, the hypersurface foliation algebra should not
change. We stress that the same hypersurface foliation algebra appears
\textit{both} in GR and higher order theories \cite{DerrHD}. It would
therefore be worth investigating the possibility of a ``seventh route
to higher derivative theories'', in analogy to the ``seventh route to
geometrodynamics'' \cite{HKT76}, since it seems that 
 the same ``seventh route'' could lead under different assumptions to
 a theory different from GR, namely to the whole class of higher
 derivative theories of gravity. 

It is instructive to count the number of degrees of freedom, which we
can do in three different ways, cf. \cite{OUP}, Sec.~4.2.3 for the
general counting procedure. The first way is based on the
original phase space and proceeds as follows. There are 32 phase space
variables --- 12 in the three-metric sector, 12 in the extrinsic
curvature sector, and eight in the lapse-shift sector. There are 10 first
class constraints --- five primary and five secondary --- for which one can
invoke 10 gauge fixing conditions. This leaves $(32-10-10)/2=6$
degrees of freedom, which agree with earlier results and with the
content of the linearized theory presented in \cite{Rieg}. The second 
way of counting is based on our unimodular-conformal configuration
variables. We have seen that scale density $a$ and trace density $\bar{K}$ are
absent from the theory. We can thus go to the subspace spanned by
$\bar{N},N^{i},\bar{h}_{ij},\bar{K}_{ij}^{\ssst\rm T}$ 
and their canonical conjugates. These are 28 degrees of freedom in
phase space. Since $\bar{P}$ and $\mathcal{Q}^{\ssst\rm W}$ were
already taken into account to expel $a$ and $\bar{K}$ from the
formalism, we have to take into account only eight constraints, leading again
to $(28-8-8)/2=6$ degrees of freedom.  

There is yet another method of counting the degrees of freedom. Let us
consider the configuration space instead of the phase space to count
the \textit{physical}, propagating, degrees of freedom. We have ten
configuration variables, five in $\bar{h}_{ij}$ and five in
$\bar{K}_{ij}^{\ssst\rm T}$. Since we do not have any restrictions on
these variables with respect to conformal transformations, there are
only four constraints left to reduce the number of variables to $10-4=6$
degrees of freedom. 

A simple thought about an alternative derivation of the Hamiltonian
analysis is of use in understanding the previous statement. Suppose we
first used unimodular-conformal variables at the unconstrained
\textit{Lagrangian} level. It would then follow from the evidence
presented so far that scale density $a$ and trace density $\bar{K}$ were
completely absent at the Lagrangian level, too.  The scale constraint
\eqref{eqn:Qw} would not be present at all, since there would be no
variables which transform under conformal transformations. This is
clear once one takes into account \eqref{eqn:pa}, which with
\eqref{eqn:Qw} implies that the Lagrange multiplier $\lambda$ is
unnecessary. The counting then takes place in such a way that it does
not include arbitrary variables. Namely, there would be four vanishing
momenta, implying that the lapse density and shift are arbitrary. We
are then left with ten variables ($\bar{h}_{ij}$ and
$\bar{K}_{ij}^{\ssst\rm T}$) which are not constrained by any
scale constraint, since they are conformally
invariant. Reparametrization invariance (Hamiltonian and momentum
constraints) constrains the ten configuration variables to $10-4=6$
degrees of freedom, which agrees with the above counting. 

Note that the linearized theory possesses ghost degrees of freedom,
which cannot be deduced from the constraint analysis by a simple
counting of degrees of freedom. One can see, however, that the
Hamiltonian is linear in momenta, hence seeming unstable (unbounded from
below), and one may expect to deal either with negative energies or with
nonunitarity (in quantum theory). When referring to such an
instability, one must, however, keep in mind that we have a
Hamiltonian {\em constraint}. This means that any negative part
arising from linearity in momenta is compensated by a
corresponding positive term, so the negative term is tamed and most
likely harmless. In fact, already in GR, part of the kinetic term in
the Hamiltonian constraint (the one connected with intrinsic time) is
negative definite, without causing instabilities; in fact, this term
is needed for good reasons \cite{GK94}.  

\subsection{Conformal symmetry and the generator of conformal
  transformation.} 

It was claimed in \cite{Blw,DerrHD,Kluson2014} that the constraint
arising from $\dot{\bar{P}}\approx 0$, that is, the conformal
constraint [in our case, this is the scale constraint \eqref{eqn:Qw}],
is the generator of conformal transformations on the original
variables. Kluson {\em et al}. \cite{Kluson2014}, for example, arrived at a
conformal constraint of the following form,\footnote{We use here the
  original variables.}  
\begin{equation}
\label{eqn:QKluson}
\mathcal{Q}:=2h_{ij}p^{ij}+K_{ij}P^{ij}\approx 0,
\end{equation}
where $P^{ij}$ was not decomposed and is thus not traceless. They claimed, in particular, that $\mathcal{Q}$ generates a conformal transformation of $K_{ij}$,
\begin{equation}
\left\lbrace K_{ij},\mathcal{Q}[\omega]\right\rbrace = \omega K_{ij}\,.
\end{equation}
This is, however, in contradiction to the actual conformal transformation of $K_{ij}$ given by \eqref{eqn:infconfK}, because the above Poisson bracket cannot produce the inhomogeneous term  containing the derivatives of $\omega$. It would be a correct transformation if one had split $K_{ij}$ into traceless and trace parts and noticed that the term $KP\approx 0$ in \eqref{eqn:QKluson} already vanishes weakly, so that only
\begin{equation}
\mathcal{Q}=2h_{ij}p^{ij}+K_{ij}^{\ssst\rm T}P^{ij\ssst\rm T}\approx 0
\end{equation}
should be demanded as a constraint.\footnote{Note that the first term is just (twice) the trace of the momentum, which is related to $p_{a}$.} In this case, the action of $\mathcal{Q}$ could be considered as generating a conformal transformation, but of $K_{ij}^{\ssst\rm T}$ only,
\begin{equation}
\left\lbrace K_{ij}^{\ssst\rm T},\mathcal{Q}[\omega]\right\rbrace = \omega K_{ij}^{\ssst\rm T}\,.
\end{equation}
A problem would still remain for the trace $K$, because the Poisson
bracket of $K$ with $\mathcal{Q}$ would be trivially 0, in
contradiction with \eqref{eqn:confKtr}. Thus, the action of
$\mathcal{Q}$ alone cannot fully recover the transformation law for
all variables---a piece of information is missing, and it has to
involve derivatives of the conformal transformation parameter,
$\del_{t}\omega\,,\,\del_{i}\omega$. 

This apparent contradictory nature of the conformal transformation
was first noticed by Irakleidou et al. \cite{ILP}, who used
Castellani's algorithm \cite{Cast} [more precisely, the
Anderson-Bergmann-Castellani (ABC) algorithm \cite{AB,Cast}] to
derive the generator of gauge transformations and proved, among other
things, that its correct form contains both the primary constraint $P$
(corresponding to $\bar{P}$ in our case) and the secondary constraint
arising from the consistency condition $\dot{P}\approx 0$. The
generator also contains the  momentum constraints in order to include
conformal transformations of the lapse. Another way to identify this
contradiction is to consider the second piece of information, which is
the lack of physical interpretation of the isolated constraint
$P\approx 0$, as noticed by \cite{Kluson2014}, namely, the
counterintuitive conformal-like nature of transformations on $p^{ij}$
and $K_{ij}$, 
\begin{equation}
\label{old_trafo}
\left\lbrace K_{ij},P[\epsilon]\right\rbrace = \epsilon h_{ij}\,,\qquad\left\lbrace p^{ij},P[\epsilon]\right\rbrace = -\epsilon P^{ij}.
\end{equation}
Checking whether the trace of the first equation in \eqref{old_trafo} yields the correct conformal transformation for $K$, one finds $\left\lbrace K,P[\epsilon]\right\rbrace = 3\epsilon $, 
which is obviously a contradiction. The resolution of the problem can be found in \cite{PSS1997,PSS2000}, where the ABC algorithm for diffeomorphism and internal gauge transformations is carefully analyzed. Here, we use a simplified version of their results, as emphasized by Pitts \cite{Pitts} (see also the references therein), which states that primary and secondary constraints have to ``work together in a tuned sum'' in order to give the correct generator of gauge transformations. Roughly speaking, they are put together in such a way that the transformation parameter of the primary constraint $\epsilon$ and the one corresponding to the secondary constraint $\omega$ have to be related in a specific way, namely
\begin{equation}
\epsilon=-\dot{\omega}.
\end{equation}
It is easy to see that in the present case this is not enough, since
the spatial derivatives to complete the Lie derivative of $\omega$ are
missing. A rigorous treatment following \cite{PSS1997,PSS2000} could
solve the problem, and we refer the reader to these papers for further
insight. Here, we extend Pitts' reasoning to the spatial variations
and make the following identification: 
\begin{equation}
\label{eqn:Bext}
\epsilon=-\mathcal{L}_{n}\omega.
\end{equation}
Putting the constraints $P$ and $\mathcal{Q}$ together, we propose the following expression as the generator for conformal transformations:
\begin{equation}
\label{eqn:Gwold}
G_{\omega}[\omega,\dot{\omega}]:=\int {\rm d}^3 x\left(\mathcal{Q}\cdot\omega+P\cdot\mathcal{L}_{n}\omega\right).
\end{equation}
To check if this gives the correct result, we calculate the Poisson
bracket of $K_{ij}$ with $G$ to find
\begin{equation}
\left\lbrace K_{ij},G_{\omega}[\omega,\dot{\omega}]\right\rbrace =\omega K_{ij} + h_{ij} \mathcal{L}_{n}\omega,
\end{equation}
which exactly agrees with \eqref{eqn:infconfK}.

However, as stated in \cite{ILP}, the generator of conformal
transformations should include the transformation of the lapse, in
accordance with \eqref{eqn:confN}, for which the simple ``sum tuning''
does not work. The authors of \cite{ILP} used the ABC recipe to derive
the following generator: 
\begin{equation}
\label{eqn:ILPconfG}
G_{\omega}[\omega,\dot{\omega}]=\int {\rm d}^3 x\left( \frac{1}{N}\dot{\omega}P+\omega\left(\mathcal{Q}+Np_{\ssst N}+\mathcal{L}_{\vec{N}}\frac{P}{N}\right)\right),
\end{equation}
which generates correctly the conformal transformations \eqref{eqn:conf3met}--\eqref{eqn:confKtr}.

Returning to our case, only $a$, $p_{a}$, and $\bar{K}$ change under a conformal transformation, due to the choice of unimodular-conformal variables ($\bar{P}$ does not change since it vanishes). Therefore, we expect that the appropriate generator of the conformal transformation reproduces \eqref{eqn:infconf}. Leaving a rigorous derivation via the ABC algorithm for another time, we take the primary constraint $\bar{P}$ and the secondary constraint $\mathcal{Q}^{\ssst \rm W}$ and form a tuned sum. Namely, if we let $\bar{P}[\epsilon]+\mathcal{Q}^{\ssst \rm W}[\omega]$ act on $\bar{K}$ and $a$, we get
\begin{equation}
\left\lbrace \bar{K},\bar{P}[\epsilon] + \mathcal{Q}^{\ssst \rm W}[\omega]\right\rbrace=\epsilon,\quad \left\lbrace a,\bar{P}[\epsilon] + \mathcal{Q}^{\ssst \rm W}[\omega]\right\rbrace=a\omega.
\end{equation}
By comparing with \eqref{eqn:infconf}, it can be seen that one should
make the identification \eqref{eqn:Bext} (with $n$ replaced by
$\bar{n}$) to define the generator of
the conformal transformation as follows: 
\begin{align}
\label{eqn:confG}
G_{\omega}^{\ssst\rm W}[\omega,\dot{\omega}]&=\intx \left(\omega\cdot\mathcal{Q}^{\ssst\rm W}+\bar{P}\cdot\mathcal{L}_{\bar{n}}\omega\right)\nonumber\\
&=\intx \left(\omega a p_{a}+\bar{P}\mathcal{L}_{\bar{n}}\omega\right).
\end{align}
Comparing with the result (84) from \cite{ILP}, one can see that the lapse momentum term is missing [apart from the last term differing by a boundary term from \eqref{eqn:confG}]. This is because $\bar{N}$ is conformally invariant. One can now check that the conformal variations of all variables vanish trivially, except for the scale density $a$ and the trace density $\bar{K}$, for which one gets
\begin{align}
\label{eqn:confGaK}
\delta_{\omega}a &=\left\lbrace a, G_{\omega}^{\ssst\rm W}[\omega,\dot{\omega}]\right\rbrace=\left\lbrace a, \int{\rm d}^3 x \,\omega\, a p_{a}\right\rbrace=\omega a\,,\nonumber\\
\delta_{\omega}\bar{K} &=\left\lbrace \bar{K}, G_{\omega}^{\ssst\rm W}[\omega,\dot{\omega}]\right\rbrace=\left\lbrace \bar{K}, \int{\rm d}^3 x \,\bar{P}\mathcal{L}_{\bar{n}}\omega\right\rbrace= \mathcal{L}_{\bar{n}}\omega,
\end{align}
in agreement with \eqref{eqn:infconf}. It is now clear why one should
use $\mathcal{Q}^{\ssst\rm W}=ap_{a}$ in \eqref{eqn:QW} instead of
$\mathcal{Q}^{\ssst\rm W}=p_{a}$ only; $\delta_{\omega}a$ has to be
proportional to $a$, and our generator $G_{\omega}^{\ssst\rm
  W}[\omega,\dot{\omega}]$ incorporates this demand, giving the
correct transformation laws for $a$ and $\bar{K}$. Using
\eqref{eqn:confinvobj}, one can easily check that our results are in
agreement with \cite{ILP}. In this sense, \eqref{eqn:confGaK} along
with \eqref{eqn:confinvobj} can be considered as ``building blocks''
of conformal transformations in the $3+1$ formalism. The result is
that \textit{all} other secondary constraints are \textit{conformally
  invariant}, which improves the previous result \cite{Blw,Kluson2014}
that the Hamiltonian constraint is conformally
\textit{covariant}. Moreover, the absence of $a$ and $\bar{K}$ from
the theory means that these variables do not need to be
gauge fixed. We thus have exploited the full power of the conformal
symmetry present in the Weyl theory. 

To summarize, we have constructed here the generator of gauge
transformations by putting primary and secondary constraints,
$\bar{P}$ and $\mathcal{Q}^{\ssst\rm W}$, into a tuned sum, using
\eqref{eqn:Bext}, to derive the correct form of the conformal
transformation generator in unimodular-conformal variables. It should
be emphasized that this identification (with $n$ replaced by
$\bar{n}$) is not general and only works
in the present case. Our treatment is possible because we work with a
conformally invariant lapse density $\bar{N}$, and because the gauge
transformations are of a particular simple form; for a derivation of
all gauge generators in GR and the Weyl-tensor theory in original
variables, see \cite{Cast} and \cite{ILP}, respectively. A rigorous
derivation of gauge generators can be found in
\cite{PSS1997,PSS2000}. We believe that the method presented there gives different results for the explicit form of the generators in
the original and the unimodular-conformal variables, due to their
different behavior under conformal transformations.  

\subsection{Weyl-Hamilton-Jacobi functional}

In 1962, Peres has shown \cite{Peres} that the Hamiltonian constraint
of GR can be written as a Hamilton-Jacobi equation, by introducing a
functional $S^{\ssst\rm E}[h_{ij}]$ as its solution, such that the ADM
momentum is defined as $p_{\ssst\rm ADM}^{ij}=\delta S^{\ssst\rm
  E}/\delta h_{ij}$. The resulting equation became known as
the ``Einstein-Hamilton-Jacobi equation'' (EHJ) and reads 
\begin{equation}
\label{eqn:EHJ}
\frac{2\kappa}{\sqrt{h}}\mathcal{G}_{ikjl}\frac{\delta S^{\ssst\rm E}}{\delta h_{ij}}\frac{\delta S^{\ssst\rm E}}{\delta h_{kl}}-\frac{\sqrt{h}}{2\kappa}\,^{\ssst (3)}\! R=0,
\end{equation}
where
\begin{equation}
\mathcal{G}_{ikjl}=\frac{1}{2}\left(h_{ik}h_{jl}+h_{il}h_{jk}-h_{ij}h_{kl}\right)
\end{equation}
is the inverse of the DeWitt metric \cite{Witt67,GK94},
\begin{equation}
\mathcal{G}^{ikjl}=\frac{1}{2}\left(h^{ik}h^{jl}+h^{il}h^{jk}-2h^{ij}h^{kl}\right).
\end{equation}
Gerlach \cite{Ger} then showed that the EHJ is equivalent to all ten Einstein field equations. In analogy to \cite{Peres}, we introduce a functional 
\begin{equation}
S^{\ssst\rm W}=S^{\ssst\rm W}[\bar{h}_{ij},a,\bar{K}_{ij}^{\ssst\rm T},\bar{K}],
\end{equation}
which is defined on the \textit{full} configuration space in the unimodular-conformal basis, such that the conjugate momenta $\bar{p}^{ij},p_{a},\bar{P}^{ij}$, and $\bar{P}$ are given by
\begin{align}
\label{eqn:HJmomC}
&\bar{p}^{ij}=\frac{\delta S^{\ssst\rm W}}{\delta \bar{h}_{ij}}\,,\quad\bar{P}^{ij}=\frac{\delta S^{\ssst\rm W}}{\delta \bar{K}_{ij}^{\ssst\rm T}}\,,\\
\label{eqn:HJmomS}
& p_{a}=\frac{\delta S^{\ssst\rm W}}{\delta a}\,,\quad\bar{P}=\frac{\delta S^{\ssst\rm W}}{\delta \bar{K}}\,.
\end{align} 
The Hamiltonian constraint turns into what we call the
Weyl-Hamilton-Jacobi (WHJ) equation \cite{KN17a}, 
\begin{align}
\label{eqn:WHJ}
\mathcal{H}^{\ssst\rm W}_{\bot}&=-\frac{1}{2\alpha_{\ssst\rm W}}\bar{h}_{ik}\bar{h}_{jl}\frac{\delta S^{\ssst\rm W}}{\delta \bar{K}_{ij}^{\ssst\rm T}}\frac{\delta S^{\ssst\rm W}}{\delta \bar{K}_{kl}^{\ssst\rm T}}+\left(\,^{\ssst (3)}\! \bar{R}_{ij}^{\ssst\rm T}+\del_{i}\bar{D}_{j}\right)\frac{\delta S^{\ssst\rm W}}{\delta \bar{K}_{ij}^{\ssst\rm T}}\nonumber\\[6pt]
&\quad+2\bar{K}_{ij}^{\ssst\rm T}\frac{\delta S^{\ssst\rm W}}{\delta \bar{h}_{ij}}-\alpha_{\ssst\rm W}\bar{C}_{ijk}^2\approx 0\ .
\end{align}
The momenta \eqref{eqn:HJmomC} and \eqref{eqn:HJmomS} should be replaced in all other constraints, too. One may, in analogy to \cite{Ger}, try to prove that the WHJ equation is equivalent to the Bach equations (equations of motion for the Weyl action); see, for example, \cite{DS} for some solutions to this equation. Here, however, we only focus on the functional $S^{\ssst\rm W}[\bar{h}_{ij},a,\bar{K}_{ij}^{\ssst\rm T},\bar{K}]$ itself and discuss its invariance under gauge transformations.

The Hamiltonian and momentum constraints reflect the fact that $S^{\ssst\rm W}$ is invariant under time reparametrizations and three-diffeomorphisms, respectively. The generator of conformal transformations suggested by \eqref{eqn:confG} should imply conformal invariance of the functional $S^{\ssst\rm W}$, as we show now.

Conformal variations of the functional $S^{\ssst\rm W}$ only act along the $\delta a$ and $\delta \bar{K}$ directions in the unimodular-conformal basis of the configuration space [using \eqref{eqn:infconf}],
\begin{align}
\label{eqn:delS}
\delta_{\omega}S^{\ssst\rm W}&=\intx \left(\delta_{\omega}a\frac{\delta S^{\ssst\rm W}}{\delta a}+\delta_{\omega}\bar{K}\frac{\delta S^{\ssst\rm W}}{\delta \bar{K}}\right)\nonumber\\
&=\intx \left(\omega a\frac{\delta S^{\ssst\rm W}}{\delta a}+\mathcal{L}_{\bar{n}}\omega\frac{\delta S^{\ssst\rm W}}{\delta \bar{K}}\right)\,.
\end{align}
On the one hand, one expects to have $\delta_{\omega}S^{\ssst\rm W}=0$. On the other hand, we have seen from the discussion so far that the correct generator of conformal transformation is given by \eqref{eqn:confG}, and we have 
\begin{equation}
\label{eqn:confGcnst}
G^{\ssst\rm W}_{\omega}[\omega,\dot{\omega}]=0
\end{equation}
on the constraint surface. Now one can easily see that
\eqref{eqn:confG} coincides with \eqref{eqn:delS} when the momenta are
expressed in terms of $S^{\ssst\rm W}$. Thus, one concludes that  
\begin{equation}
\label{eqn:GSzero}
G^{\ssst\rm W}_{\omega}[\omega,\dot{\omega}]=0\,\,\wedge \,\, G^{\ssst\rm W}_{\omega}[\omega,\dot{\omega}]=\delta_{\omega}S^{\ssst\rm W}\,\,\Rightarrow\,\,\delta_{\omega}S^{\ssst\rm W}=0,
\end{equation}
that is, that the evaluation of $G_{\omega}[\omega,\dot{\omega}]$ on the constraint surface expresses the invariance of $S^{\ssst\rm W}$ under conformal transformation. It follows that $S^{\ssst\rm W}$ is independent of scale density $a$ and trace density $\bar{K}$,
\begin{equation}
\label{eqn:W-HJS}
S^{\ssst\rm W}=S^{\ssst\rm W}[\bar{h}_{ij},\bar{K}_{ij}^{\ssst\rm T}]\,,
\end{equation}
as expected. One might ask which relation holds instead of \eqref{eqn:GSzero} and what its meaning is if one attempts to construct such expressions in a theory with second class constraints, as a consequence of conformal symmetry breaking. This is discussed in the following section.

\section{Adding the Einstein-Hilbert term: Weyl-Einstein gravity}
\label{ClassWE}

\subsection{Hamiltonian formulation}

Supplementing the Weyl-tensor action \eqref{eqn:W-action} by the EH
term using the expression \eqref{eqn:Rdec} for the Ricci scalar, the
velocity $\mathcal{L}_{n}K$ is explicitly introduced into the
action,\footnote{We ignore the total divergence coming from
  $D_{i}D_{j}N/N$ .} 
\begin{align}
\label{eqn:WE31c}
\tilde{S}^{\ssst \rm WE}=\int{\!\rm d}t\,{\rm d}^3x
  &\,N\sqrt{h}\Biggl\lbrace-\frac{\alpha_{\ssst\rm W}}{2}
    C_{ij}^{\ssst\rm T}C^{ij\ssst\rm T}+\alpha_{\ssst\rm W}
    C_{ijk}^{2}\nonumber\\ 
&+\frac{1}{2\kappa}\left(2\mathcal{L}_{n}K+\,^{\ssst (3)}\! R+K_{ij}K^{ij}+K^2\right)\nonumber\\
&-\lambda^{ij}\left(2K_{ij}-\mathcal{L}_{n}h_{ij}\right)\Biggr\rbrace. 
\end{align}
As a consequence, the constraint $P\approx 0$ no longer holds, which
is an explicit signal for the conformal symmetry breaking. Instead of
using \eqref{eqn:Rdec} in \eqref{eqn:WE31c}, one could alternatively
perform a partial integration of the term $\mathcal{L}_{n}K$ to arrive
at the usual expression for the EH action and the well-known boundary
term [this is equivalent to using \eqref{eqn:Rdec1}] 
\begin{equation}
\label{eqn:GHY}
\sigma\frac{1}{\kappa}\int {\rm d}^{3}x\sqrt{h}K,
\end{equation}
where $\sigma = -1 $ or $+1$ for the timelike or spacelike boundary, respectively.

In GR, it is customary to supplement the EH action with the negative
of the above term in order to get rid of the second time derivative of
the metric (which translates to the first time derivative of $K$) and
provide second order equations of motion upon variation with respect
to the metric.\footnote{This was already noticed by Einstein
  \cite{Einstein16}.} 
 Here, we are however dealing with a higher derivative theory, whose defining feature is exactly to have second order time derivatives of the metric which cannot be reduced and whose equations of motion are fourth order. Note that the boundary term \eqref{eqn:GHY} vanishes upon variation, since $K_{ij}$ is also required besides $h_{ij}$
  to be fixed on the boundary. Therefore, a partial integration of the
  $K$-velocity term is not needed in squared curvature actions
  supplemented by the EH term, and it is harmless; see
  \cite{Kluson2014} and references therein. We choose to use the
  latter (i.e. the partially integrated) version of the EH action,
  because the EH term then leaves the constraint $\bar{P}\approx 0$
  intact. This choice does not make the breaking of conformal symmetry
  obvious at first glance, but it simplifies the resulting Poisson
  brackets. In which way the choice of (not) using the boundary term
  affects the quantum theory is left to be investigated elsewhere. 

The simplified action takes the following form:
\begin{align}
\label{eqn:WE31}
S^{\ssst \rm WE}=\int{\!\rm d}t\,{\rm d}^3x &\,N\sqrt{h}\Biggl\lbrace-\frac{\alpha_{\ssst\rm W}}{2} C_{ij}^{\ssst\rm T}C^{ij\ssst\rm T}+\alpha_{\ssst\rm W} C_{ijk}^{2}\nonumber\\
&+\frac{1}{2\kappa}\left(\,^{\ssst (3)}\! R+K_{ij}K^{ij}-K^2\right)\nonumber\\
&-\lambda^{ij}\left(2K_{ij}-\mathcal{L}_{n}h_{ij}\right)\Biggr\rbrace\,.
\end{align}
Using unimodular-conformal variables, the conjugate momenta for the above action are the same as in \eqref{eqn:pNbar}--\eqref{eqn:pKbar}. The total Hamiltonian is given by
\begin{align}
\label{eqn:WEHHam}
H^{\ssst\rm WE}=&\int{\rm d}^3 x\Biggl\lbrace \bar{N}\Biggl[-\frac{\bar{h}_{ik}\bar{h}_{jl}\bar{P}^{ij}\bar{P}^{kl}}{2\alpha_{\ssst\rm W}}+\left(\,^{\ssst (3)}\! \bar{R}_{ij}^{\ssst\rm T}+\del_{i}\bar{D}_{j}\right)\bar{P}^{ij}\nonumber\\[6pt]
&+2\bar{K}_{ij}^{\ssst\rm T}\bar{p}^{ij}+a\bar{K}p_{a}-\alpha_{\ssst\rm W}\bar{C}_{ijk}^2\nonumber\\[6pt]
&-\frac{1}{2\kappa}a^4\left(\,^{\ssst\rm (3)}\! R+a^{-2}\bar{K}_{ij}^{{\ssst\rm T }2}-6\,a^{-2}\bar{K}^{2}\right)\Biggr]\nonumber\\[6pt]
&+N^{i}\Biggl[-2D_{k}
\left(\bar{h}_{ij}\bar{p}^{jk}\right)-\frac{1}{3}D_{i}\left(a\,p_{a}\right)\nonumber\\[6pt]
&-2D_{k}\left(\bar{K}_{ij}^{\ssst\rm T}\bar{P}^{jk}\right)+D_{i}\bar{K}_{jk}^{\ssst\rm T}
\bar{P}^{jk}\Biggr]\nonumber\\[6pt]
&+\lambda_{\ssst\bar{N}}p_{\ssst\bar{N}}+\lambda_{i}p^{i}+\lambda_{\ssst\bar{P}}\bar{P}\Biggr\rbrace+H_{\rm surf},
\end{align}
where $\bar{K}_{ij}^{{\ssst\rm T }2}\equiv\bar{K}_{ij}^{\ssst\rm
  T}\bar{h}^{ia}\bar{h}^{jb}\bar{K}_{ab}^{\ssst\rm T}$, but $\,^{\ssst
  (3)}\! R$ is not yet decomposed. Let us again first discuss the
conservation of the $\bar{P}\approx 0$ constraint, which leads to the
requirement 
\begin{align}
\label{eqn:PcnstrWEH}
\dot{\bar{P}}=&-\frac{\delta H}{\delta \bar{K}}=-\bar{N}\left(ap_{a}+\frac{6a^2}{\kappa}\bar{K}\right)\approx 0,\nonumber\\
&\qquad\Rightarrow\quad \mathcal{Q}^{\ssst\rm WE}:= ap_{a}+\frac{6a^2}{\kappa}\bar{K}\approx 0,
\end{align}
where we have included $a$ in the definition of $\mathcal{Q}^{\ssst\rm
  WE}$ for reasons similar to the ones in the pure Weyl case. The
``spoiled'' conformal constraint turns out to form a pair of
second class constraints with $\bar{P}\approx 0$, since the
constraints do not commute, 
\begin{equation}
\label{eqn:WE-PQcomm}
\left\lbrace \bar{P}({\mathbf x}),\,\mathcal{Q}^{\ssst\rm
    WE}({\mathbf y})\right\rbrace=-\frac{6a^2}{\kappa}\delta({\mathbf
  x}-{\mathbf y}).  
\end{equation}
It is precisely the trace density $\bar{K}$ (which is proportional to
$\dot{a}$) that is responsible for this,  due to the expression 
\begin{equation}
\label{eqn:WE-confbreakQ}
\mathcal{Q}^{\ssst\rm E}:=\frac{6a^2}{\kappa}\bar{K},
\end{equation}
which depends on $\kappa$. Observe that for $\kappa^{-1}\rightarrow 0$ (which would correspond to a high energy regime beyond the Planck scale where the Weyl-tensor term dominates) the constraints become first class and the conformal symmetry is restored. It is now clear that this behavior obviously needs to be treated within a quantum theory of gravity, and that is exactly what we are going to do in a follow-up paper \cite{KN17b}.
In addition, it is clear that $\mathcal{Q}^{\ssst\rm WE}$ is not
automatically conserved in time, since the consistency condition
produces terms which are not weakly 0. Demanding the time
derivative to be 0,
\begin{align}
\label{eqn:WE-Plagmult}
\dot{\mathcal{Q}}^{\ssst\rm WE}&=\left\lbrace\mathcal{Q}^{\ssst\rm WE},H^{\ssst\rm WE}\right\rbrace \nonumber\\
&= \frac{\bar{N}}{2\kappa}\Biggl(2a^{2}\,^{\ssst (3)}\! \bar{R}-4a\del_{i}\left(\bar{h}^{ij}\del_{j}a\right)+12a^2\bar{K}^2\nonumber\\
&\quad+2a^2\bar{K}_{ij}^{{\ssst\rm T }2}\Biggr)+\frac{4a^2}{\kappa}\bar{K}D_{i}N^{i}+\frac{6a^2}{\kappa}\lambda_{\bar{P}}\approx 0,
\end{align}
determines the Lagrange multiplier $\lambda_{\bar{P}}$,
\begin{align}
\lambda_{\bar{P}}&=-\frac{\bar{N}}{6a^2}\left(a^{2}\,^{\ssst (3)}\! \bar{R}-2a\del_{i}\left(\bar{h}^{ij}\del_{j}a\right)+6a^2\bar{K}^2\right.\nonumber\\
&\quad\left.+a^2\bar{K}_{ij}^{{\ssst\rm T
  }2}\right)-\frac{2}{3}\bar{K}D_{i}N^{i},
\end{align}
effectively determining $\dot{\bar{K}}$, which is undetermined in the
pure W theory. Note that our result for this Lagrange multiplier
differs from the one obtained in \cite{Kluson2014}, because we use
unimodular-conformal canonical variables, which expel the variable
$\bar{P}$ from the constraints. 

We do not insert this value for the Lagrange multiplier into
\eqref{eqn:WEHHam}, but calculate instead Dirac brackets. 
Let us first derive from \eqref{eqn:WEHHam} the Hamiltonian and the
momentum constraints. Conservation of the constraint
$p_{\ssst\bar{N}}\approx 0$ in time leads to 
\begin{align}
\label{eqn:pNdot-Hc}
\dot{p}_{\ssst\bar{N}}&=\left\lbrace p_{\ssst\bar{N}}, H^{\ssst\rm
                        WE}\right\rbrace\nonumber\\ 
&=-\frac{\bar{h}_{ik}\bar{h}_{jl}\bar{P}^{ij}\bar{P}^{kl}}{2\alpha_{\ssst\rm
  W}}+\left(\,^{\ssst (3)}\! \bar{R}_{ij}^{\ssst\rm
  T}+\del_{i}\bar{D}_{j}\right)\bar{P}^{ij}\nonumber\\[6pt] 
&\quad+2\bar{K}_{ij}^{\ssst\rm T}\bar{p}^{ij}+a\bar{K}p_{a}-\alpha_{\ssst\rm W}\bar{C}_{ijk}^2\nonumber\\[6pt]
&-\frac{1}{2\kappa}a^4\left(\,^{\ssst (3)}\!
  R+a^{-2}\bar{K}_{ij}^{{\ssst\rm
  T}2}-6\,a^{-2}\bar{K}^{2}\right)\approx 0. 
\end{align}
We already know from \eqref{eqn:PcnstrWEH} that $a p_{a}+6a^2\bar{K}/\kappa\approx 0$; if we use this relation in \eqref{eqn:pNdot-Hc}, we arrive at the following expression for the Hamiltonian constraint:
\begin{align}
\label{eqn:HamWE}
\mathcal{H}^{\ssst\rm WE}_{\bot}&=-\frac{\bar{h}_{ik}\bar{h}_{jl}\bar{P}^{ij}\bar{P}^{kl}}{2\alpha_{\ssst\rm W}}+\left(\,^{\ssst (3)}\! \bar{R}_{ij}^{\ssst\rm T}+\del_{i}\bar{D}_{j}\right)\bar{P}^{ij}\nonumber\\[6pt]
&\quad+2\bar{K}_{ij}^{\ssst\rm T}\bar{p}^{ij}-\alpha_{\ssst\rm W}\bar{C}_{ijk}^2\nonumber\\[6pt]
&\quad-\frac{1}{2\kappa}a^4\left(\,^{\ssst (3)}\! R+a^{-2}\bar{K}_{ij}^{{\ssst\rm T}2}+6\,a^{-2}\bar{K}^{2}\right)\nonumber\\[6pt]
&=\mathcal{H}^{\ssst\rm W}_{\bot}-\frac{1}{2\kappa}a^4\left(\,^{\ssst (3)}\! R+a^{-2}\bar{K}_{ij}^{{\ssst\rm T}2}+6\,a^{-2}\bar{K}^{2}\right)\approx 0,
\end{align}
where $\mathcal{H}^{\ssst\rm W}_{\bot}$ is the Hamiltonian constraint
of the pure W theory,  which is given in \eqref{eqn:HamWcf}. It is
clear that due to the term 
\begin{equation}
\label{eqn:WE-confbreakH}
\mathcal{H}_{\bot}^{\ssst\rm E}:= -\frac{1}{2\kappa}a^4\left(\,^{\ssst (3)}\! R+a^{-2}\bar{K}_{ij}^{{\ssst\rm T}2}+6\,a^{-2}\bar{K}^{2}\right),
\end{equation}
the Hamiltonian constraint is now explicitly {\em not} invariant under
conformal transformations. The momentum constraint in the WE theory
contains an additional term $D_{i}\left(a\,p_{a}\right)$ which is
scale dependent, as is the case with the scale constraint. The
constraints read 
\begin{align}
\label{eqn:MomWE}
\mathcal{H}^{\ssst\rm WE}_{i}&=-2\del_{k}\left(
\bar{h}_{ij}\bar{p}^{jk}\right)+\del_{i}\bar{h}_{jk}\bar{p}^{jk}-\frac{1}{3}D_{i}\left(a\,p_{a}\right),\nonumber\\[6pt] 
&\quad-2\del_{k}\left(\bar{K}_{ij}^{\ssst\rm
  T}\bar{P}^{jk}\right)+\del_{i}\bar{K}_{jk}^{\ssst\rm
  T}\bar{P}^{jk}\approx 0\\[12pt] 
\label{eqn:QWE}
\mathcal{Q}^{\ssst\rm WE}&=ap_{a}+\frac{6a^2}{\kappa}\bar{K}\approx 0.
\end{align}
The theory is now manifestly not conformally invariant. The use of our
unimodular-conformal variables makes this clear by revealing the
scale density- and trace density-dependent terms which are responsible for
conformal symmetry breaking in the WE theory. 

We can now formulate the total Hamiltonian as a linear combination of all constraints:
\begin{align}
\label{eqn:fintotHWE}
H^{\ssst\rm WE}=\int{\rm d}^3 x\Biggl\lbrace &\bar{N}\mathcal{H}_{\bot}^{\ssst\rm WE}+N^{i}\mathcal{H}^{\ssst\rm WE}_{i}+\left(\bar{N}\bar{K}\right)\mathcal{Q}^{\ssst\rm WE}\nonumber\\[6pt]
&+\lambda_{\bar{N}}p_{\bar{N}}+\lambda_{i}p^{i}+\lambda_{\bar{P}}\bar{P}\Biggr\rbrace\,,
\end{align}
where $\bar{N}\cdot\bar{K}$ can be interpreted as the Lagrange multiplier for the $\mathcal{Q}^{\ssst\rm WE}$ constraint.\footnote{Compare this with the result from \cite{Kluson2014} where $\mathcal{Q}^{\ssst\rm WE}$ was added \textit{by hand} with a Lagrange multiplier $\lambda_{Q}$, which of course led the authors to the conclusion that $\lambda_{Q}=0$. The reason why they got such a result is that there is no need to add the secondary second class constraint by hand---it is already present in the total Hamiltonian, and one simply needs to identify it from the appropriate consistency condition.} If we had, instead, eliminated $a\,p_{a}$ from \eqref{eqn:MomWE} using \eqref{eqn:QWE} (in the spirit of what we did in the pure Weyl case), the momentum constraint would read $\mathcal{H}^{\ssst\rm W}_{i}+2a^2D_{i}\bar{K}/\kappa \approx 0$, and the Lagrangian multiplier to $\mathcal{Q}^{\ssst\rm WE}$ would have the same form as in the pure Weyl case. It is unclear to us which procedure is the more appropriate one.

What can we say about the constraint algebra in the WE theory?
Comparing with the algebra \eqref{eqn:AlgHH}--\eqref{eqn:AlgPQ} in the
pure Weyl theory, we expect that the first three Poisson brackets
remain unchanged because the hypersurface foliation algebra should be
preserved (the Weyl and EH terms in the action fulfil this
symmetry separately). Concerning the other constraints, we observe
that because of \eqref{eqn:WE-confbreakQ} we now have
\be
\label{eqn:AlgPQwe}
\left\lbrace \bar{P}[\epsilon],\mathcal{Q}^{\ssst\rm
    WE}[\omega]\right\rbrace =-\frac{6}{\kappa}\intx
\,\epsilon\,\omega\, a^2, 
\ee
and the commutation of the constraints is thus spoiled. This is, of course, a consequence of the conformal symmetry breaking.

Since we are dealing here with a system that has both first and second
class constraints, leading to one determined Lagrange multiplier,
the Poisson brackets should be replaced by Dirac brackets. For a
general function $F({\mathbf x})$ and $G({\mathbf x})$, the Dirac
bracket reads \cite{Dir}  
\begin{align}
\label{eqn:DB}
&\left\lbrace F({\mathbf x}),G({\mathbf
  y})\right\rbrace_{D}=\left\lbrace F({\mathbf x}),G({\mathbf
  y})\right\rbrace\nonumber\\ 
&-\int {\rm d}^{3}z\,{\rm d}^{3}z'\left\lbrace F({\mathbf
  x}),\phi_{A}({\mathbf z})\right\rbrace\mathcal{M}^{AB}\left\lbrace
  \phi_{B}({\mathbf z}'),G({\mathbf y})\right\rbrace,
\end{align}
where the sum is understood as running over the second class
constraints here labelled by $A,B=(1,2)$ with
$\phi_{1}({\mathbf z})=\bar{P}({\mathbf z})$ and
$\phi_{2}({\mathbf z})=(ap_{a}+\frac{6}{\kappa}a^{2}\bar{K})({\mathbf z})$, and
$\mathcal{M}^{AB}$ is the inverse matrix to 
\begin{align}
\mathcal{M}_{AB}&=
\begin{pmatrix}
\left\lbrace \phi_{1}({\mathbf z}),\phi_{1}({\mathbf z}')\right\rbrace& \left\lbrace \phi_{1}({\mathbf z}),\phi_{2}({\mathbf z}')\right\rbrace\\[6pt]
\left\lbrace \phi_{2}({\mathbf z}),\phi_{1}({\mathbf z}')\right\rbrace& \left\lbrace \phi_{2}({\mathbf z}),\phi_{2}({\mathbf z}')\right\rbrace
\end{pmatrix}\nonumber\\[6pt]
&=
-\frac{6a^2}{\kappa}\begin{pmatrix}
0 & 1\\
-1 & 0
\end{pmatrix},
\end{align}
so
\begin{equation}
\mathcal{M}^{AB}=\frac{\kappa}{6a^{2}}
\begin{pmatrix}
0 & 1\\
-1 & 0
\end{pmatrix}.
\end{equation}

The form of $\phi_{1}$ and $\phi_{2}$ is such that only brackets
between $\bar{K}$ and $a,p_{a},\bar{P}$ are modified. This concerns
the following three brackets:
\begin{align}
\label{eqn:W-DB3}
\left\lbrace \bar{K},\bar{P}\right\rbrace_{D}&=0,\nonumber\\ 
\left\lbrace \bar{K},a\right\rbrace_{D}&=\frac{\kappa}{6a},\\
\left\lbrace \bar{K},p_{a}\right\rbrace_{D}&\approx-\frac{\bar{K}}{a}\nonumber,
\end{align}
where in the last bracket we have used the constraint \eqref{eqn:QWE}. 
Because of the use of unimodular-conformal variables, 
the resulting Dirac brackets are of a much simpler form than the ones
derived in \cite{Kluson2014}. This clearly shows which part of the
configuration space is affected by conformal symmetry breaking. The
above brackets show that $\bar{K}$ can no longer be treated as a
``true'' canonical variable---it is determined by other variables in
the theory. Counting the degrees of freedom proceeds as follows: we
have 12 degrees of freedom to start, including $a$ and $\bar{K}$, and
four first class constraints, which gives eight degrees of freedom. We also
have one pair of second class constraints, which eliminates one more
degree of freedom, so we end up with a total of seven physical degrees of
freedom, which agrees with previous results \cite{Kluson2014}. 

\subsection{Hamilton-Jacobi functional and the generator of nongauge
  conformal transformations} 
\label{HJWE}

In the same way as in the pure Weyl case, a functional $S^{\ssst\rm
  WE}[\bar{h}_{ij},a,\bar{K}^{\ssst\rm T}_{ij},\bar{K}]$ can be
introduced to express the momenta as functional derivatives. The
Hamilton-Jacobi equation that follows from \eqref{eqn:HamWE} is then
given by 
\begin{align}
\label{eqn:HJWE}
&-\frac{1}{2\alpha_{\ssst\rm W}}\bar{h}_{ik}\bar{h}_{jl}\frac{\delta
  S^{\ssst\rm WE}}{\delta\bar{K}^{\ssst\rm T}_{ij}}\frac{\delta
  S^{\ssst\rm WE}}{\delta\bar{K}^{\ssst\rm T}_{kl}}+\left(\,^{\ssst
  (3)}\! \bar{R}_{ij}^{\ssst\rm
  T}+\del_{i}\bar{D}_{j}\right)\frac{\delta S^{\ssst\rm
  WE}}{\delta\bar{K}^{\ssst\rm T}_{ij}}\nonumber\\[6pt] 
&+2\bar{K}_{ij}^{\ssst\rm T}\frac{\delta S^{\ssst\rm WE}}{\delta\bar{h}_{ij}}-\alpha_{\ssst\rm W}\bar{C}_{ijk}^2\nonumber\\[6pt]
&-\frac{1}{2\kappa}a^4\left(\,^{\ssst (3)}\! 
  R+a^{-2}\bar{K}_{ij}^{{\ssst\rm T
  }2}+6\,a^{-2}\bar{K}^{2}\right)=0. 
\end{align}
One may refer to this equation as the ``Weyl-Einstein-Hamilton-Jacobi
equation'' (WEHJ). We 
see that the WEHJ is qualitatively different from both the EHJ and the
WHJ, owing to the different structure of the Hamiltonian
constraint. The signature of symmetry breaking is the explicit
presence of scale density $a$ and trace density $\bar{K}$, making $S^{\ssst\rm
  WE}$ noninvariant under conformal transformation. Thus, it would
seem that $S^{\ssst\rm WE}$ now remains to live on the full
configuration space. We recognize, however, from the Dirac brackets
\eqref{eqn:W-DB3} that $\bar{K}$ is not to be treated as a
configuration space variable at
all---it is fixed by the EH term, and it is not a true degree of
freedom. This is clarified below. 

At first glance, the fact that
\begin{equation}
\label{eqn:WE-HJp0}
\bar{P}\approx 0\quad\Rightarrow\quad \frac{\delta S^{\ssst\rm WE}}{\delta \bar{K}}=0
\end{equation}
seems to be contradictory with the explicit appearance of $\bar{K}$ in
\eqref{eqn:HJWE}. How can this be? Equation \eqref{eqn:WE-HJp0} should
be interpreted merely as a statement that $S^{\ssst\rm WE}$ should
\textit{functionally} not depend on $\bar{K}$ \textit{as a
  configuration variable}. Based only on \eqref{eqn:WE-HJp0}, nothing
more can be said---additional information is required. The secondary
constraint $\mathcal{Q}^{\ssst\rm WE}$ when evaluated with the HJ functional
provides the second piece of information, 
\begin{equation}
\label{eqn:WE-HJQ}
\mathcal{Q}^{\ssst\rm WE}\approx 0\quad\Rightarrow\quad a\frac{\delta
  S^{\ssst\rm WE}}{\delta a}=-\frac{6a^2}{\kappa}\bar{K}. 
\end{equation}
Since $\delta S^{\ssst\rm WE}/\delta a$ no longer vanishes, the
interpretation of \eqref{eqn:WE-HJQ} is that the variation of $a$ is no
longer arbitrary, which is consistent with the observation from the Dirac brackets
\eqref{eqn:W-DB3} that $a$ is no longer an arbitrary variable. This
variation seems to be fixed by $a$ and $\bar{K}$ due to the presence of
the EH term [note the coupling $1/\kappa$ in \eqref{eqn:WE-HJQ}].

 The final piece of information is provided by the fact that
the primary-secondary pair of constraints $\bar{P}$ and
$\mathcal{Q}^{\ssst\rm WE}$ are of \textit{second class}. This
effectively means that the price for the nonarbitrary variation of $a$ is
that $\bar{K}$ becomes fixed through \eqref{eqn:WE-HJQ}. Closer inspection of
\eqref{eqn:WE-HJQ} reveals that this is just the trace of the ADM
momentum $p^{ij}_{\ssst\rm ADM}=\sqrt{h}(K^{ij}-h^{ij}K)/2\kappa$ of
the pure EH-theory as expressed in terms of the
unimodular-conformal variables, see in this connection \cite{BNdice}.
This suggests that conformal symmetry breaking
gives rise to the \textit{time evolution of scale density}
$a$ \textit{by determining it through $\bar{K}$}. Therefore, one can
say that conformal symmetry breaking by the EH term (or any
term explicitly dependent on $\bar{K}$) ``downgrades'' $\bar{K}$ from
a configuration space variable to a function with a fixed form.  

What does all this imply for the generator of gauge conformal
transformations? Since the involved constraints are of second class, such
a generator cannot be defined in the WE theory, at least not using the
ABC algorithm which determines only gauge generators using
\textit{first} class constraints. However, a closer look at the
action \eqref{eqn:confGaK} of the generator in W theory suggests that
the form of the generator in the W theory may be used 
as a general definition independent of the context of that theory
because the conformal variations $\delta_{\omega} a = \omega a$ and
$\delta_{\omega} \bar{K} = \mathcal{L}_{\bar{n}}\omega$ are determined
only by the definitions of the respective variables (the only variables
which change under conformal transformation) and the specific form of
the conformal transformation for the metric. One may thus use
\begin{equation}
\label{eqn:GSgeneral}
G_{\omega}[\omega,\dot{\omega}]=\intx \left(\omega \cdot a\,
  p_{a}+\bar{P}\cdot\mathcal{L}_{\bar{n}}\omega \right) 
\end{equation}
as an ansatz for the \textit{generator of conformal
  transformations}\footnote{Again, we warn the reader not to be misled by
  terminology---the generator refers to transformations
  \eqref{eqn:cftrans}, which are essentially local Weyl rescalings of
  fields.} \textit{in the Hamiltonian formulation}, meaning that such a
generator may be defined independently of a particular theory and studied in any
theory expressed in the $3+1$ formalism in unimodular-conformal
canonical variables (hence the superscript ``W'' is omitted). With
this in mind, we can write \eqref{eqn:GSzero} again, with the difference
that in the WE theory it does not vanish, 
\begin{equation}
\label{eqn:GSWEn0}
G_{\omega}[\omega,\dot{\omega}]=\delta_{\omega}S^{\ssst\rm WE}\neq 0\,.
\end{equation}
The above statement is not a postulate, but follows directly when \eqref{eqn:GSgeneral} is evaluated on the constraint hypersurface of the WE theory with \eqref{eqn:WE-HJp0} and \eqref{eqn:WE-HJQ}, that is,
\begin{align}
\label{eqn:delSWE}
\delta_{\omega}S^{\ssst\rm WE}&=\intx \left(\omega a\frac{\delta S^{\ssst\rm WE}}{\delta a}+\mathcal{L}_{\bar{n}}\omega\frac{\delta S^{\ssst\rm WE}}{\delta \bar{K}}\right)\nonumber\\[6pt]
&=-\frac{6}{\kappa}\intx\,\omega\, a^2 \bar{K}\neq 0\,,
\end{align}
from which \eqref{eqn:GSWEn0} reads
\begin{equation}
\label{eqn:GSWEres}
G_{\omega}[\omega,\dot{\omega}]=-\frac{6}{\kappa}\intx\,\omega\,a^2\bar{K}\,.
\end{equation}
This is an important result. It means that evaluating the generator
of conformal transformation \eqref{eqn:GSgeneral} on the constraint
hypersurface in the configuration space of the WE theory fails to give
a vanishing variation of the Hamilton-Jacobi functional, which implies
that the conformal symmetry in the WE theory is broken. Since $S^{\ssst\rm WE}$
functionally does not depend on $\bar{K}$ (see \eqref{eqn:WE-HJp0}),
and variation of $S^{\ssst\rm WE}$ with respect to scale density $a$ is
nonarbitrary and relates the trace density $\bar{K}$ to the momentum
$p_{a}$ [see \eqref{eqn:WE-HJQ}], the HJ functional of the WE theory
depends on scale density, that is, 
\begin{equation}
\label{eqn:WE-Sa-dep}
S^{\ssst\rm WE}=S^{\ssst\rm WE}[\bar{h}_{ij},a,\bar{K}^{\ssst\rm T}_{ij}]\,.
\end{equation}
Since scale density is the variable affected by conformal transformation, the HJ functional of the WE theory is \textit{not} conformally invariant, in contrast to the HJ functional of the pure W theory [see \eqref{eqn:W-HJS}].

To summarize, a theory has a symmetry under conformal (rescaling)
transformation only if the generator of conformal transformations
\eqref{eqn:GSgeneral} vanishes on the constraint hypersurface in the
configuration space of the theory. We think that a similar conclusion may
be generalized to hold for other symmetries. 

We conclude this subsection with a remark on the generator of
conformal transformations for fields. As mentioned earlier, in the
presently discussed case of the WE theory the ABC algorithm cannot be
applied to find the generator of gauge conformal transformation, since
constraints are second class and a conformal transformation is not a
gauge one. However, suppose we naively use the prescription of a tuned
sum in this case, we might think of putting the
primary-secondary pair of second class constraints  into the following
expression: 
\begin{align}
\label{eqn:WE-G2nd}
\tilde{G}_{\omega}[\omega,\dot{\omega}]&\equiv\intx\left(\omega\cdot\mathcal{Q}^{\ssst\rm WE}+\bar{P}\cdot\mathcal{L}_{\bar{n}}\omega\right)\nonumber\\
&=\intx\left(\omega\cdot\left(ap_{a}+\frac{6a^2}{\kappa}\bar{K}\right)+\bar{P}\cdot\mathcal{L}_{\bar{n}}\omega\right). 
\end{align}
This object does, of course, not have the meaning of a gauge generator of
conformal transformations, but it is interesting to note that it 
still produces correct conformal transformations for the scale density $a$ and
the trace density $\bar{K}$ (note that this may not be true in a
setting more general than the WE theory). Moreover, we note that the
requirement for this object to vanish on the constraint hypersurface
in the configuration space, 
\begin{equation}
\label{eqn:WE-G2ndzero}
\tilde{G}_{\omega}[\omega,\dot{\omega}]\stackrel{!}{=}0\,,
\end{equation}
implies the result found earlier in \eqref{eqn:GSWEres}, which is obvious if one rewrites \eqref{eqn:WE-G2nd} and \eqref{eqn:WE-G2ndzero} as
 \begin{align}
\label{eqn:WE-G2ndzero1}
\tilde{G}_{\omega}[\omega,\dot{\omega}]=G_{\omega}[\omega,\dot{\omega}]+\frac{6}{\kappa}\intx\,\omega \,a^2\bar{K}=0.
\end{align}
It would then be of interest to investigate whether it is possible to
formulate such an object $\tilde{G}$ in \textit{any} constrained
system containing second class constraints, and whether they are
always connected with the breaking of some symmetry, that is, whether one
can always impose the requirement \eqref{eqn:WE-G2ndzero}, but this is
beyond the scope of the present work. 

We are now ready to include matter into the theory and discuss
symmetries and generators of transformations of the full WE theory
with matter.

\section{Adding matter: Weyl-Einstein gravity with nonminimally
  coupled scalar field} 
\label{ClassWEm}

\subsection{Hamiltonian formulation}

For a more realistic picture, it is necessary to extend the WE theory by including a nongravitational (matter) sector.
 We consider the Lagrangian of a nonminimally coupled scalar field
 (including the case of minimal coupling as a special case) given by 
\begin{align}
\label{eqn:nm-action}
\mathcal{L}^{\varphi}&=-\frac{1}{2}\sqrt{-g}\Biggl[g^{\mu\nu}\del_{\mu}\varphi\del_{\nu}\varphi+\xi R\varphi^2+V(\varphi)\Biggr]\nonumber\\
&=\frac{1}{2}N\sqrt{h}\Biggl[\left(n^{\mu}\del_{\mu}\varphi\right)^2+\xi\left(\frac{2}{3}K^{2}-2\nabla_{\mu}\left(n^{\mu}K\right)\right)\varphi^2\nonumber\\
&\quad-h^{ij}\del_{i}\varphi\,\del_{j}\varphi-\xi\left( \,^{\ssst (3)}R-\frac{2}{N}D^{i}D_{i}N\right)\varphi^2\nonumber\\
&\quad-\xi K_{ij}^{\ssst\rm T}K^{ij\ssst\rm T}\varphi^2-V(\varphi)\Biggr],
\end{align}
where the nonminimal coupling constant $\xi$ is dimensionless, and where we have used \eqref{eqn:Rdec1} simply because it is easier to work with than with \eqref{eqn:Rdec}. It is known that for $\xi=1/6$ the above action is invariant under conformal transformation \eqref{eqn:cftrans} if the scalar field is transformed as
\begin{equation}
\label{eqn:nm-phiscaled}
\varphi\rightarrow \tilde{\varphi} =\Omega^{-1}\varphi\,,
\end{equation}
along with the metric, and if the potential term is either 0 or $\beta\varphi^4$, with $\beta$ being a dimensionless coupling. If the scalar field is massive, the conformal symmetry is broken, and if $\xi=0$ we have the usual minimally coupled scalar field, for which the conformal symmetry is also lost. The essential features of our investigation can be seen already for the case $V(\varphi)=0$ to which we restrict ourselves from now on. It is clear that, depending on the value of $\xi$, we will either have only first or first and second class constraints, but in both cases we will be able to define a generator of conformal transformation.

The $3+1$ decomposition of the above action can be found, for example, in \cite{Kluson2014,KiefNM} and the references therein. It is convenient to \textit{rescale} the scalar field to $\varphi\rightarrow a \varphi$, such that the resulting variable does not transform under a conformal transformation. We can thus introduce a new scalar density by
\begin{equation}
\label{eqn:nm-chi}
\chi:= a \varphi,
\end{equation}
which is of scale weight\footnote{Defined in Appendix \ref{AppRicciBar}.} $w_{a}=1$. It is also possible to define this scalar density as
 $\chi:=a^{6\xi}\varphi$ for which it was shown in \cite{KiefNM} for vanishing shift
 that the interaction term $\dot{\varphi}K$
is eliminated from the $\varphi$-Lagrangian, cf. Eq. (2.3) there. 
However, we choose to use \eqref{eqn:nm-chi} because in that case the scaling of the scalar field $\varphi$ with $a^{-1}$ reflects its physical dimension (inverse length)
 and $\chi$ is dimensionless (see Appendix \ref{appdim}), while a more general discussion is possible if the scaling is independent of $\xi$. 
Rescaling the scalar
field in this way is also important in cosmological perturbation
theory; see, for example, \cite{BKK16}. This choice of scaling is
particularly suggestive because it
compensates in \eqref{eqn:nm-chi} the necessary conformal rescaling of $\varphi$ in
\eqref{eqn:nm-phiscaled}, resulting in a conformally invariant scalar
density variable $\chi$. 

Employing unimodular-conformal variables, along with
\eqref{eqn:nm-chi}, the Lagrangian can be decomposed in the following way: 
\begin{align}
\label{eqn:nm-Lagdec}
\mathcal{L}^{\chi}&=\frac{1}{2}\bar{N}\Bigg[\left(\bar{n}^{\mu}\del_{\mu}\chi-(1-6\xi)\bar{K}\chi-\frac{\del_{i}N^{i}}{3\bar{N}}\chi\right)^2\nonumber\\[6pt]
&\quad +6\xi(1-6\xi)\bar{K}^2\chi^2\nonumber\\[12pt]
&\quad -\xi\,^{\ssst (3)}\!\bar{R}\chi^2 +4\xi\chi\del_{i}\left(\bar{h}^{ij}\del_{j}\chi\right)+\left(4\xi-1\right)\bar{h}^{ij}\del_{i}\chi\del_{j}\chi\nonumber\\[6pt]
&\quad -\xi\bar{K}_{ij}^{{\ssst\rm T}2}\,\chi^2+(1-6\xi)S(a)  \Bigg]+\xi\del B,
\end{align} 
where $\bar{K}_{ij}^{{\ssst\rm T}2}\equiv\bar{K}_{ij}^{\ssst\rm
  T}\bar{h}^{ia}\bar{h}^{jb}\bar{K}_{ab}^{\ssst\rm T}$, as before, and 
\begin{align}
\label{eqn:totDiv}
\del B&:=
\del_{i}\left(\bar{h}^{ij}\left(\del_{j}\bar{N}\chi^2-\bar{N}\del_{j}\chi^2\right)\right)+3\del_{i}\left(\bar{N}\bar{h}^{ij}\del_{j}\log a\chi^2\right)\nonumber\\[6pt]
&\quad-3\del_{\mu}\left(\bar{N}\bar{n}^{\mu}\bar{K}\chi^2\right)\ .
\end{align}
is a collection of total divergences resulting from several applications of the Leibniz rule to extract the lapse density from the derivatives. The third line in the above decomposed Lagrangian \eqref{eqn:nm-Lagdec}, along with the last term $(1-6\xi)S(a)$, results from the combination $h^{ij}\del_{i}\varphi\,\del_{j}\varphi-\xi\left( \,^{\ssst (3)}R-2D^{i}D_{i}N/N\right)\varphi^2$ in the original Lagrangian \eqref{eqn:nm-action}; the term $S(a;\xi)$ is given by
\begin{align}
\label{eqn:Sscales}
S(a)=\bar{h}^{ij}\left(\del_{i}\log a\,\del_{j}\chi^2-\del_{i}\log a\,\del_{j}\log a\,\chi^2\right)
\end{align}
and is a collection of all scale-dependent terms arising from the decomposition of $\,^{\ssst (3)}\! R$ [see \eqref{eqn:RicciScalDec}] and the rest of the terms in the third line of \eqref{eqn:nm-action}. The first and the second lines in \eqref{eqn:nm-Lagdec} result from the first line of \eqref{eqn:nm-action} and from $\xi\,^{\ssst (3)}\! R\varphi^2$.

It is obvious that all terms in the Lagrangian \eqref{eqn:nm-Lagdec}
are \textit{separately} conformally invariant, {\em except} for three terms proportional to
$\left(1-6\xi\right)$. These are the only terms which depend either on $a$ or $\bar{K}$. It is clear that the latter three terms have
something in common: they vanish for $\xi=1/6$, that is, for
conformal coupling. Hence, if these terms were absent, the Lagrangian
of the nonminimally coupled scalar field for $\xi=1/6$ would be
\textit{conformally invariant term by term}, 
\begin{align}
\label{eqn:nm-LagConf}
&\mathcal{L}^{\chi}=\frac{1}{2}\bar{N}\Bigg[\left(\bar{n}^{\mu}\del_{\mu}\chi-\frac{\del_{i}N^{i}}{3\bar{N}}\chi\right)^2-\frac{1}{6}\Big(\,^{\ssst (3)}\!\bar{R}+\bar{K}_{ij}^{{\ssst\rm T}2}\Big)\chi^2\nonumber\\[6pt]
&\quad  -\frac{1}{3}\left(2\chi\del_{i}\left(\bar{h}^{ij}\del_{j}\chi\right)-\bar{h}^{ij}\del_{i}\chi\del_{j}\chi\right) \Bigg]+\frac{1}{6}\del B\ ,
\end{align}
up to a total divergence $\del B$. It can then be concluded that \textit{the conformal
  symmetry requires the absence of scale density} $a$ \textit{and trace
  density} $\bar{K}$ \textit{from the action, up to a surface
  integral}. In other words, absence of $a$ and $\bar{K}$ is the
fingerprint of conformal invariance.\footnote{Recall again that we talk
  about the local rescaling of fields, not conformal coordinate
  transformations.} It may now be understood in the context of the
$3+1$ decomposition why the nonminimal coupling term $\xi R\varphi^2$
is necessary for the conformal invariance of the scalar field action:
it serves to eliminate $a$ and $\bar{K}$ from those terms in
$g^{\mu\nu}\del_{\mu}\varphi\del_{\nu}\varphi$ for $\xi=1/6$ that
contain scales and traces. Note that this
cancellation was made possible by the particular choice of scaling
\eqref{eqn:nm-chi} for the scalar field (see, for example, Appendix~D
in \cite{Wald}) which reflects its physical dimension of inverse length. 

After this preparation, we can address the Hamiltonian formulation.
The canonical momentum conjugate to $\chi$ is given by
\begin{equation}
\label{eqn:chimom}
p_{\chi}=\frac{\del\mathcal{L}^{\chi}}{\del\dot{\chi}}=\bar{n}^{\mu}\del_{\mu}\chi-(1-6\xi)\bar{K}\chi-\frac{\del_{i}N^{i}}{3\bar{N}}\chi\ ,
\end{equation}
and the total \textit{matter}\footnote{Here, we ignore the gravitational variables in the Legendre transformation because we want to focus on the expressions arising solely from matter.} Hamiltonian follows to read
\begin{align}
\label{eqn:nm-Ham}
H^{\chi}&=\intx\left(\dot{\chi}p_{\chi}-\mathcal{L}^{\chi}\right)\nonumber\\
&=\intx \left\lbrace\bar{N}\mathcal{H}^{\chi}_{\bot}+N^{i}\mathcal{H}^{\chi}_{i}\right\rbrace+H_{\rm surf}^{\chi},
\end{align}
where
\begin{align}
\label{eqn:nm-HamC}
\mathcal{H}^{\chi}_{\bot}&:=\frac{1}{2}\Bigg[p_{\chi}^2+\xi\left(\bar{R}+\bar{K}_{ij}^{{\ssst\rm T}2}\,\right)\chi^2\nonumber\\[6pt]
&\quad -4\xi\chi\del_{i}\left(\bar{h}^{ij}\del_{j}\chi\right)+\left(1-4\xi\right)\bar{h}^{ij}\del_{i}\chi\del_{j}\chi\nonumber\\[6pt]
&\quad +(1-6\xi)\Big(2\bar{K}\chi\, p_{\chi}-6\xi\bar{K}^2\chi^2-S(a)\Big)  \Bigg],\\[6pt]
\label{eqn:nm-MomC}
\mathcal{H}^{\chi}_{i}&:= -\frac{1}{3}\left(\chi\del_{i}p_{\chi}-2\del_{i}\chi \,p_{\chi}\right),\\[6pt]
\label{eqn:nm-Hsurf}
H_{\rm surf}^{\chi}&:=\xi\intx \left[ 2\del_{i}\left(N^{i}\chi p_{\chi}\right)-\del B\right]
\end{align}
are the matter contributions to the Hamiltonian constraint, the
momentum constraints, and the surface integral, respectively. They
contribute to the Hamiltonian and momentum constraints of the
Weyl-Einstein-$\chi$ theory (WE$\chi$ for short). Note that for
$\xi=1/6$ the expressions \eqref{eqn:nm-HamC} and \eqref{eqn:nm-MomC}
are both manifestly conformally invariant, so this feature also holds
for the full Hamiltonian \eqref{eqn:nm-Ham} (up to the surface
term). For the sake of completeness, we give here the matter
contribution to the Hamiltonian constraint in the conformally
invariant case: 
\begin{align}
\label{eqn:nm-HamC1}
&\mathcal{H}^{\chi}_{\bot}\left(\xi=\frac{1}{6}\right)= \frac{1}{2}p_{\chi}^{2}+ \frac{1}{12}\biggl(\,^{\ssst (3)}\! \bar{R}+\bar{K}_{ij}^{{\ssst\rm T}2}\biggr)\chi^2\nonumber\\
&\quad-\frac{1}{6}\Biggl(4\chi\del_{i}\left(\bar{h}^{ij}\del_{j}\chi\right)-2\bar{h}^{ij}\del_{i}\chi\del_{j}\chi \Biggr).
\end{align}
This expression does not contain any scales or traces. On the other hand, note that in the case of \textit{minimal coupling} ($\xi = 0$), both $a$ and $\bar{K}$ are present, reflecting the conformally noninvariant nature of such coupling.

The reason why we can simply add the various Hamiltonians is because we treat $\bar{h}_{ij},a,\bar{K}_{ij}^{\ssst\rm T},\bar{K}$ as independent variables (below to be introduced for the full WE$\chi$ theory) and because \eqref{eqn:nm-Lagdec} does not contain the velocities $\mathcal{L}_{\bar{n}}\bar{K}$ (application of the Leibniz rule led to a boundary term). For this reason, we do not need to consider the term $\dot{\bar{K}}\bar{P}$ in the Hamiltonian \eqref{eqn:nm-Ham}. (This was also the case for the W and WEH theories above.) If, instead, we had chosen to keep the term $\mathcal{L}_{\bar{n}}\bar{K}$, there would have been a nonzero contribution to $\bar{P}$ due to the presence of $\mathcal{L}_{\bar{n}}\bar{K}\chi^2$. This would lead to a different formulation that deserves further study.

We can now put the Lagrangians of the WE and $\chi$ theories together and proceed to the Hamiltonian formulation using the definitions \eqref{eqn:pNbar}--\eqref{eqn:pKbar} and \eqref{eqn:chimom} for the momenta and replacing $\mathcal{L}^{\ssst\rm W}_{c}\rightarrow\mathcal{L}^{{\ssst\rm WE}\chi}_{c}=\mathcal{L}^{\ssst\rm WE}_{c}+\mathcal{L}^{\chi}$. The Legendre transform to the total Hamiltonian then gives
\begin{align}
\label{eqn:WEnm-Hamtot}
H^{{\ssst\rm WE}\chi}&=\intx\,\Biggl\lbrace \dot{\bar{h}}_{ij}\bar{p}^{ij}+\dot{a}p_{a}+\dot{\bar{K}}_{ij}^{\ssst\rm T}\bar{P}^{ij}+\dot{\chi}p_{\chi}-\mathcal{L}_{c}^{{\ssst\rm WE}\chi}\nonumber\\
&\qquad+\lambda_{\ssst\bar{N}}p_{\ssst\bar{N}}+\lambda_{i}p^{i}+\lambda_{\ssst\bar{P}}\bar{P}\Biggr\rbrace,
\end{align}
from which one can determine the following secondary constraints
\begin{align}
\label{eqn:WEnm-HamC}
\mathcal{H}^{{\ssst\rm WE}\chi}_{\bot}&:=\mathcal{H}^{\ssst\rm W}_{\bot}+\mathcal{H}^{\ssst\rm E}_{\bot}+\mathcal{H}^{\chi}_{\bot}\approx 0,\\[6pt]
\label{eqn:WEnm-MomC}
\mathcal{H}^{{\ssst\rm WE}\chi}_{i}&:=\mathcal{H}^{\ssst\rm WE}_{i}+\mathcal{H}^{\chi}_{i}\approx 0,\\[6pt]
\label{eqn:WEnm-QC}
\mathcal{Q}^{{\ssst\rm WE}\chi}&:=\mathcal{Q}^{\ssst\rm W}+\mathcal{Q}^{\ssst\rm E}+\mathcal{Q}^{\chi}\approx 0.
\end{align}
It is obvious from this formulation of the constraints that
 $\mathcal{H}^{\ssst\rm W}_{\bot}$ comes from \eqref{eqn:HamWcf}, $\mathcal{H}^{\ssst\rm E}_{\bot}$ comes from \eqref{eqn:WE-confbreakH}, $\mathcal{H}^{\chi}_{\bot}$ comes from \eqref{eqn:nm-HamC}, and that the momentum constraints consist of \eqref{eqn:MomWE} and \eqref{eqn:nm-MomC}; the expressions $\mathcal{Q}^{\ssst\rm W}$ and $\mathcal{Q}^{\ssst\rm E}$ are defined in \eqref{eqn:QW} and \eqref{eqn:WE-confbreakQ}, respectively, while $\mathcal{Q}^{\chi}$ originates from the matter part of the theory and reads
\begin{equation}
\label{eqn:WEnm-QChi}
\mathcal{Q}^{\chi}:=(1-6\xi)\chi\, p_{\chi} -6\xi\left(1-6\xi\right)\bar{K}\chi^2\,.
\end{equation}

The explicit form of the constraints is determined as follows. It will prove convenient to introduce 
\begin{equation}
\label{eqn:WEnm-kappa}
\frac{1}{\tilde{\kappa}}:=\frac{1}{\kappa}-\xi\left(1-6\xi\right)\frac{\chi^2}{a^2}\,,
\end{equation}
which could be interpreted as coupling between $\chi$ and the scale  and trace density $a,\bar{K}$, see below. The consistency condition for $\bar{P}\approx 0$ is such that it requires the expression $\mathcal{Q}^{{\ssst\rm WE}\chi}=ap_{a}+(1-6\xi)\chi\, p_{\chi}+6a^2\bar{K}/\tilde{\kappa}$ to vanish. It is now not possible to use this constraint in order to completely eliminate $\bar{K}$ from the remaining constraints, in contrast to the pure Weyl case. Namely, the consistency condition for $p_{\bar{N}}$ leads to\footnote{After identifying $\mathcal{Q}^{{\ssst\rm WE}\chi}\approx 0$, similarly to the step from \eqref{eqn:pNdot-Hc} to \eqref{eqn:HamWE} in the WE case.} a Hamiltonian constraint which depends on $a$ and $\bar{K}$; the constraints read
\begin{align}
\label{eqn:WEnm-HamCall}
\mathcal{H}^{{\ssst\rm WE}\chi}_{\bot}&=\mathcal{H}^{\ssst\rm W}_{\bot}+\frac{1}{2}p_{\chi}^2
+ 3\xi^2\left(\,^{\ssst (3)}\bar{R}+\bar{K}_{ij}^{{\ssst\rm T}2}\right)\chi^2\nonumber\\[6pt]
&-\frac{1}{2}\biggl(4\xi\chi\del_{i}\left(\bar{h}^{ij}\del_{j}\chi\right)-\left(1-4\xi\right)\bar{h}^{ij}\del_{i}\chi\del_{j}\chi\biggr)\nonumber\\[6pt]
&-\frac{a^2}{2\tilde{\kappa}}\biggl(\,^{\ssst (3)}\bar{R}+\bar{K}_{ij}^{{\ssst\rm T}2}+6\bar{K}^2\biggr)\nonumber\\[6pt]
&+\frac{a^2}{\tilde{\kappa}}
	\Bigl(
	2\del_{i}\left(\bar{h}^{ij}\del_{j}\log a\right)+\bar{h}^{ij}\del_{i}\log a \,\del_{j}\log a\Bigr)\nonumber\\[6pt]
	&+(1-6\xi)\Bigl(
	4\xi\del_{i}\left(\bar{h}^{ij}\del_{j}\log a\right)\chi^2-\bar{h}^{ij}\del_{i}\log a \,\del_{j}\chi^2\nonumber\\[6pt]
	&\quad+(1+2\xi)\bar{h}^{ij}\del_{i}\log a \,\del_{j}\log a\,\chi^2
	\Bigr)\approx 0,\\[12pt]
\label{eqn:WEnm-MomCall}
\mathcal{H}^{{\ssst\rm WE}\chi}_{i}&=-2\del_{k}\left(
\bar{h}_{ij}\bar{p}^{jk}\right)+\del_{i}\bar{h}_{jk}\bar{p}^{jk}-\frac{1}{3}D_{i}\left(a\,p_{a}\right)\nonumber\\[6pt]
&\quad-2\del_{k}\left(\bar{K}_{ij}^{\ssst\rm T}\bar{P}^{jk}\right)+\del_{i}\bar{K}_{jk}^{\ssst\rm T}\bar{P}^{jk}\nonumber\\[6pt]
&\quad-\frac{1}{3}\left(\chi\del_{i}p_{\chi}-2\del_{i}\chi \,p_{\chi}\right)\approx 0\\[12pt]
\label{eqn:WEnm-QCall}
\mathcal{Q}^{{\ssst\rm WE}\chi}&=ap_{a}+(1-6\xi)\chi\, p_{\chi}+\frac{6a^2}{\tilde{\kappa}}\bar{K}\approx 0\ .
\end{align}
Note the appearance of $1/\tilde{\kappa}$ in two terms in
\eqref{eqn:WEnm-HamCall}. One should keep in mind that
$1/\tilde{\kappa}$ is not a constant, that is,
$\tilde{\kappa}=\tilde{\kappa}(a,\chi)$ and that it depends parametrically on $\xi$.

We choose to work with $\tilde{\kappa}$ instead of $\kappa$ because
one can then easily recognize the features of the Hamiltonian
constraint in terms of the couplings $\xi$ and $1/\tilde{\kappa}$ and
see more clearly the contrast between the conformal and nonconformal
versions of the theory. For example, $1/\tilde{\kappa}=0$ eliminates
the trace density $\bar{K}$ from all the constraints and also
eliminates the scale part of the Ricci scalar, while $\xi=1/6$
eliminates the coupling of $\chi$ to the scale density $a$ and the trace density $\bar{K}$. Let us therefore
now have a closer look at these couplings. 

We can distinguish between vanishing and nonvanishing $1/\tilde{\kappa}$. This reflects the commutation properties between the constraints,
\begin{equation}
\label{eqn:WEnm-PQcomm}
\left\lbrace \bar{P},\,\mathcal{Q}^{{\ssst\rm WE}\chi}\right\rbrace=-\frac{6a^2}{\tilde{\kappa}}\,.
\end{equation}
This commutator can vanish ($1/\tilde{\kappa}=0$) in the following three cases:
\begin{itemize}
\item[\textbf{1.}] $\frac{1}{\kappa}=0$ and $\xi=\frac{1}{6}$,
\item[\textbf{2.}] $\frac{1}{\kappa}=0$ and $\xi=0$,
\item[\textbf{3.}] $\chi^2=\chi_{c}^2 := a^{2}\left(\kappa\xi (1-6\xi)\right)^{-1}$.
\end{itemize}
It is important to realize that the commutator vanishes because $1/\tilde{\kappa}=0$ eliminates $\bar{K}^2$ from the total Hamiltonian, making all constraints independent of $\bar{K}$. We also note that $\kappa\to\infty$ is the strong-coupling limit $G\to\infty$. 

The constraints are first class only in the first case; this corresponds to the conformal version  $\xi=1/6$ and $1/\kappa=0$. The former turns $\tilde{\kappa}\rightarrow\kappa$ and eliminates the coupling of $\chi$ with scale density $a$, while the latter excludes the EH contribution. Then one is left with $\mathcal{H}^{{\ssst\rm WE}\chi}_{\bot}=\mathcal{H}^{\ssst\rm W}_{\bot}+\mathcal{H}^{\chi}_{\bot}(\xi=1/6)$, see \eqref{eqn:nm-HamC1}, that is, a scalar density field conformally coupled to the Weyl gravity. In this case the constraints are first class. 

In the second case, we are dealing with a minimally coupled scalar field in Weyl-squared gravity, namely $\xi=0$ eliminates the nonminimal coupling. One would expect that the conformal symmetry is broken due to conformal noninvariance of the minimally coupled scalar field. However, constraint analysis needs to be investigated further.
Namely, the secondary constraint \eqref{eqn:WEnm-QCall} simplifies to $\mathcal{Q}^{{\ssst\rm W}\chi}=ap_{a}+\chi\, p_{\chi}$, and one needs to check again its consistency condition. Several further constraints appear, since term $\chi\, p_{\chi}$ and the absence of $\bar{K}$ from this constraint prevent the Dirac algorithm from terminating until several further steps. We do not 
investigate this further here, but we emphasize that this is an interesting and
simple case for studying conformally noninvariant scalar fields. 

The third case is a very interesting one. It says that if the scalar
density field adopts a certain critical value $\chi_{c}$, then the
contribution of the trace density $\bar{K}$ from the EH part is
canceled, along with some of the spatial derivatives of the scale density, but
the scale density is still present and the theory is not conformally
invariant. Thus, one expects again a second class system with further
constraints, a matter into which we do not go here. This critical
value was already met in \cite{KiefNM} [see there Eq. (2.12)] and
turned out to be the limiting case between an indefinite and a
positive definite kinetic term in the Hamiltonian constraint.

If $1/\tilde{\kappa}\neq 0$ and $\xi\neq 0,\, 1/6$, we have a
nonconformal scalar density nonminimally coupled to WE gravity---the most general case. The consistency requirement for
$\mathcal{Q}^{{\ssst\rm WE}\chi}$ gives a nontrivial result from
which, in this most general case, the Lagrange multiplier
$\lambda_{\bar{P}}$ is determined from part of the second and part
of the third term in 
\begin{equation}
\label{eqn:WEnm-Qdot}
\dot{\mathcal{Q}}^{{\ssst\rm WE}\chi}=\left\lbrace\mathcal{Q}^{\ssst\rm W},H^{{\ssst\rm WE}\chi}\right\rbrace+\left\lbrace\mathcal{Q}^{\ssst\rm E},H^{{\ssst\rm WE}\chi}\right\rbrace+\left\lbrace\mathcal{Q}^{\chi},H^{{\ssst\rm WE}\chi}\right\rbrace,
\end{equation}
which give a nonvanishing term proportional to $(\delta \mathcal{Q}^{{\ssst\rm WE}\chi}/\delta \bar{K})(\delta H^{{\ssst\rm WE}\chi}/\delta \bar{P})\sim\lambda_{\bar{P}}$, similarly to the WE case, see \eqref{eqn:WE-Plagmult}. The Dirac algorithm has then arrived at its end, and all the constraints are determined. We do not present the explicit expression here and only give a list of the Dirac brackets,
\begin{align}
\label{eqn:W-DB-WEx}
\left\lbrace \bar{K},\bar{P}\right\rbrace_{D}&=0,\nonumber\\
\left\lbrace \bar{K},a\right\rbrace_{D}&=\frac{\tilde{\kappa}}{6a}\,,\nonumber\\
\left\lbrace \bar{K},p_{a}\right\rbrace_{D}&\approx-\frac{\bar{K}}{a}+\left(1-6\xi\right)\frac{\tilde{\kappa}}{6a^3}\chi p_{\chi},\nonumber\\
\left\lbrace \bar{K},\chi\right\rbrace_{D}&=(1-6\xi)\frac{\tilde{\kappa}}{6a^2}\chi\nonumber\\[6pt]
\left\lbrace \bar{K},p_{\chi}\right\rbrace_{D}&=\left(1-6\xi\right)\frac{\tilde{\kappa}}{6a^2}\left(p_{\chi}+12\xi\bar{K}\chi\right)\ .
\end{align}
The first three Dirac brackets are analogous to the ones derived for the WE theory. The fourth and fifth ones are there because of the matter degree of freedom and support the conclusion that $\bar{K}$ is no longer a variable in configuration space, but is fixed. The number of degrees of freedom is now 8, seven gravitational and one matter.

The results from this subsection support the claim that a conformally
invariant theory has to be independent of scale density $a$ and trace density
$\bar{K}$.

\subsection{Hamilton-Jacobi functional and the generator of
  (non-)gauge conformal transformations} 

We saw that only for particular values of couplings $1/\kappa=0$ and
$\xi=1/6$, all constraints are first class, and only in this case
can one formulate the generator of \textit{gauge} conformal transformation
by using the ABC algorithm. However, one may nevertheless consider the
most general case with $\bar{P}$ and $\mathcal{Q}^{{\ssst\rm WE}\chi}$
being the only second class constraints, by logically extending the
discussion from Sec. \ref{HJWE}, and thereby investigating the
action of the generator of conformal transformations
\eqref{eqn:GSgeneral} in the configuration space. 

We thus focus our discussion here on the HJ functional for the
general case of nonvanishing couplings and derive its change under
conformal transformations. The configuration space is now extended by
the $\chi$ variable, so that we also have here the matter momentum
$p_{\chi}=\delta S^{{\ssst\rm WE}\chi}/\delta \chi $, with
$S^{{\ssst\rm WE}\chi}=S^{{\ssst\rm
    WE}\chi}[\bar{h}_{ij},a,\bar{K}_{ij}^{\ssst\rm T},\bar{K},\chi]$
being the HJ function of the WE$\chi$ theory. This functional is
determined by the Hamilton-Jacobi equation resulting from
\eqref{eqn:WEnm-HamCall} by substituting all the momenta. 

As in the WE theory, one realizes that $\bar{P}\approx 0$ and that the relations
\begin{equation}
\label{eqn:WEnm-HJQ}
\mathcal{Q}^{{\ssst\rm WE}\chi}\approx 0\quad\Rightarrow\quad a\frac{\delta S^{{\ssst\rm WE}\chi}}{\delta a}=-(1-6\xi)\chi\, p_{\chi}-\frac{6a^2}{\tilde{\kappa}}\bar{K}
\end{equation}
imply that $S^{{\ssst\rm WE}\chi}$ functionally does not depend on
$\bar{K}$; hence we actually have 
\begin{equation}
S^{{\ssst\rm WE}\chi}=S^{{\ssst\rm WE}\chi}[\bar{h}_{ij},a,\bar{K}_{ij}^{\ssst\rm T},\chi]\,,
\end{equation}
with $\bar{K}$ fixing the variation of the HJ functional with
respect to the scale density. The interpretation of independence of
$S^{{\ssst\rm WE}\chi}$ on $\bar{K}$ is analogous to the case of the
vacuum WE theory, and we may again say that conformal
symmetry breaking gives rise to the time evolution of scale density $a$ by determining
it through $\bar{K}$. In the present case, both the EH term and
the nonminimally coupled scalar field are responsible for
this. Therefore, $\bar{K}$ is again not a configuration variable, but
a function which fixes the variation of the HJ functional with respect
to scale density $a$. 

Since we first do not impose any restrictions on the couplings, we are
dealing with broken conformal symmetry due to the presence of both
nonminimally coupled matter field $\chi$ and the EH term. Then the
generator of conformal transformations as defined generically in
\eqref{eqn:GSgeneral} determines a nonvanishing conformal
variation of the HJ functional $\delta_{\omega}S^{{\ssst\rm WE}\chi}$
when evaluated on the constraint hypersurface, 
\begin{equation}
\label{eqn:GSWEnmn0}
G_{\omega}[\omega,\dot{\omega}]=\delta_{\omega}S^{{\ssst\rm WE}\chi}\neq 0\,,
\end{equation}
similarly to the the action of the generator in the vacuum WE theory, see \eqref{eqn:GSWEn0}. It follows from 
\begin{align}
\label{eqn:delSWEnm}
\delta_{\omega}S^{{\ssst\rm WE}\chi}&=\intx \left(\omega a\frac{\delta
                                      S^{{\ssst\rm WE}\chi}}{\delta
                                      a}+\mathcal{L}_{\bar{n}}\omega\frac{\delta
                                      S^{{\ssst\rm WE}\chi}}{\delta
                                      \bar{K}}\right)\nonumber\\[6pt] 
&=-\intx\,\omega\,\left((1-6\xi)\chi\, p_{\chi}+\frac{6a^2}{\tilde{\kappa}}  \bar{K}\right)\neq 0
\end{align}
that \eqref{eqn:GSWEnmn0} does not vanish in general, 
\begin{equation}
\label{eqn:GSWEnmres}
G_{\omega}[\omega,\dot{\omega}]=-\intx\,\omega\,\left((1-6\xi)\chi\, p_{\chi}+\frac{6a^2}{\tilde{\kappa}}  \bar{K}\right)\,,
\end{equation}
thus signaling the breaking of conformal symmetry.

This formulation of the generator of conformal transformations as
defined here assumes that the definition of such a generator is independent
of the theory in question --- and thus of the number of constraints
arising from the Dirac algorithm for the primary constraint $\bar{P}$, but
we saw in Sec. \ref{HJWE} that both the class and the number of
constraints depend on the value of the couplings. How the values of
the couplings influence the above discussion is not investigated
here. Instead, let us restrict ourselves to the case where
$1/\kappa=0$ and $\xi=1/6$, which is the case of scalar density field
conformally coupled to the Weyl gravity. The choice of couplings
implies $1/\tilde{\kappa}=0$, which makes $\bar{P}$ and
$\mathcal{Q}^{{\ssst\rm WE}\chi}=\mathcal{Q}^{\ssst\rm W}$ first class
constraints and the generator of conformal transformations vanishes on
the constraint hypersurface, that is, \eqref{eqn:GSWEnmres} becomes 
\begin{equation}
1/\kappa = 0 \,\,\wedge \,\,\xi = 1/6\,\,\Rightarrow\,\,G_{\omega}[\omega,\dot{\omega}]=\delta_{\omega}S^{{\ssst\rm WE}\chi}=0\,,
\end{equation}
which is the statement of conformal invariance of the HJ functional
$S^{{\ssst\rm WE}\chi}$. The HJ functional now loses dependence on
both scale density $a$ and trace density $\bar{K}$, that is, $S^{{\ssst\rm
    WE}\chi}\rightarrow S^{{\ssst\rm W}\chi}=S^{{\ssst\rm
    W}\chi}[\bar{h}_{ij},\bar{K}_{ij}^{\ssst\rm T},\chi]$, which
solves the WHJ with conformally coupled scalar density field. In this
case, $G_{\omega}[\omega,\dot{\omega}]$ is the generator of
\textit{gauge} conformal transformations, but since no $a$ or
$\bar{K}$ appear in the theory, its action on the Hamiltonian is
trivial. Thus, we have confirmed that in the case of nonminimally
coupled scalar field the conformal symmetry is present if the scale density
$a$ and trace density $\bar{K}$ are absent from the theory, which does
happen for a particular choice of couplings. 

Note in passing that, in analogy to \eqref{eqn:WE-G2ndzero} and
\eqref{eqn:WE-G2ndzero1}, one could define an object
$\tilde{\tilde{G}}_{\omega}[\omega,\dot{\omega}]$ such that it
satisfies 
\begin{align}
\label{eqn:WEnm-G2ndzero1}
\tilde{\tilde{G}}_{\omega}[\omega,\dot{\omega}]&=G_{\omega}[\omega,\dot{\omega}]\nonumber\\[6pt]
&\quad+\intx\,\omega\,\left((1-6\xi)\chi\, p_{\chi}+\frac{6a^2}{\tilde{\kappa}}  \bar{K}\right)=0\,,
\end{align}
which would follow from 
\begin{align}
\label{eqn:WEnm-G2ndzero}
\tilde{\tilde{G}}_{\omega}[\omega,\dot{\omega}]\stackrel{!}{=}0.
\end{align}
If such an object could be formulated via a well-defined 
procedure, Eqs. \eqref{eqn:WE-G2ndzero} and \eqref{eqn:WEnm-G2ndzero}
could be identified as a universal statement providing the
answer to the question of (non-)vanishing variation of the HJ
functional and thus the question of symmetries of an underlying
theory. We do not pursue here this discussion further. 

Collecting the results from the current and the previous two sections,
one can derive the following conclusion: 

\textit{The conformal symmetry of a theory implies that the
  corresponding Hamilton-Jacobi functional is independent of scale density $a$
  and trace density $\bar{K}$. Conversely, if the Hamilton-Jacobi
  functional does not depend on scale density $a$ and trace density $\bar{K}$,
  the theory is conformally invariant, up to a total divergence.} 

It remains to be seen how general this statement is.


\section{Conclusions}

In our paper, we have developed a Hamiltonian formalism for
Weyl-squared gravity, for Weyl-squared gravity plus Einstein-Hilbert
term (WE theory), and for  Weyl-squared gravity plus Einstein-Hilbert term plus
generally coupled scalar field (WE$\chi$ theory). The new feature compared to earlier
work is the systematic use of unimodular-conformal variables, that is,
the consistent decomposition into the scale part and conformal part for
{\em all} variables. This makes especially transparent the features of
conformal invariance for Weyl-squared gravity and of conformal
symmetry breaking in the other cases. 

We have derived and discussed all constraints and their 
algebra. In the Weyl theory, the Hamiltonian and momentum constraints
have been formulated 
exclusively in terms of scale free variables. There is thus no scale
factor $a$ appearing and, consequently, no part that corresponds to an
intrinsic time, as is the case for general relativity \cite{OUP}. 
This is especially important for the quantum theory.
An interesting question for future research concerning the constraint
algebra is whether one
can derive this algebra by a ``seventh route'' to the
geometrodynamics of higher-derivative theories, in analogy to the
seventh route for Einstein's theory \cite{HKT76}. 

 We have formulated and discussed in detail the generators of symmetry
transformations. In the pure Weyl case, all constraints are of first
class. Forming a tuned sum in the spirit of \cite{Pitts}, they together
generate conformal and diffeomorphism transformations; the individual
constraints do not generate appropriate transformations. Using
again a tuned sum in the WE and WE$\chi$ theories,
 we have argued in favor of the possibility
of introducing a generator of \textit{nongauge} conformal
transformations in these theories, where two constraints are
of second class and where conformal invariance does not exist.
 This generator is able to produce appropriate conformal transformations of the
configuration space variables $a$ and $\bar{K}$, while producing a
nonvanishing change of the Hamilton-Jacobi functional in the WE and
WE$\chi$ theories. 

The next step in our investigation of those theories is
quantization \cite{KN17b}; see also \cite{KN17a} for some preliminary
results. The classical theories are now available in a form in which the scheme
of canonical quantization \cite{OUP} can be employed in a
straightforward way. We
are led to the analogues of Wheeler--DeWitt equation and quantum
momentum constraints, but also have to take into account the new
constraints. In addition, the semiclassical approximation to those
quantum constraints needs to be investigated. The final goal is, of
course, be the application of such a theory of quantum gravity to the
early Universe and to find out whether it is of empirical relevance or
not.


\appendix
\section{A note on physical units}
\label{appdim}

We list here for convenience the physical dimensions of 
some relevant quantities. Since we choose $c=1$ throughout, all
variables can be expressed in units of mass ($M$) and length ($L$). In
order to provide the scale part of the three-metric,
$a\equiv(\sqrt{h})^{1/3}$, with unit of 
length, we choose the dimension for the three-metric components to be $L^2$
and of the inverse metric components to be $L^{-2}$. This means that
the spatial coordinates are dimensionless. We keep, however, the
dimension $L$ for time, which means that the lapse function $N$ is
dimensionless. All ``barred'' (conformally invariant) configuration
space variables (those constructed from the three-metric and the
second fundamental form) including $\chi$ are then dimensionless. 
For other relevant quantities we have the dimensions
\ben
& & [a]=L,\quad [N]=1,\quad [\bar{N}]=L^{-1}, \\
 & & [N_i]=L,\quad [N^i]=[\bar{N}^i]= L^{-1}, \\
& & [K_{ij}]=[K_{ij}^{\rm T}]=L,\quad [K^{ij}]=L^{-3},\quad [K]=L^{-1},\\
& & \alpha_{\ssst\rm W}=M\cdot L,\quad [C_{ijk}]=L,\quad [C^{ijk}]=L^{-5},\\
& & [\bar{p}^{ij}]=[\bar{P}^{ij}]=[\bar{P}]=M\cdot L,\quad [p_a]=M , \quad [\varphi]=L^{-1}.
\een
Consequently, the total Hamiltonian then has the physical dimension of
a mass, as it should.
 
\section{Conformal transformations and unimodular-conformal
  variables} 
\label{AppConfUm}

Throughout the paper we refer to the following transformation as
conformal transformation, 
\begin{align}
\label{eqn:WeylRescMet}
g_{\mu\nu}(x)\quad\rightarrow &\quad \tilde{g}_{\mu\nu}(x)=\Omega^2(x)g_{\mu\nu}(x),\\
\label{eqn:conf4det}
\sqrt{-g}\quad\rightarrow &\quad \sqrt{-\tilde{g}}=\Omega^4 \sqrt{-g},
\end{align}
emphasizing that this is not a transformation of coordinates, but of
the four-metric itself. It is also referred to as local
dilatations, or local rescaling, or Weyl rescaling. It is
easy to understand from \eqref{eqn:g31mat} that the above conformal
transformation in four dimensions induces conformal transformations
for the components of the $3+1$ decomposed metric, namely, 
\begin{align}
\label{eqn:conf3met}
h_{ij}\quad\rightarrow &\quad \tilde{h}_{ij}=\Omega^2 h_{ij},\\
\label{eqn:conf3det}
\sqrt{h}\quad\rightarrow &\quad \sqrt{\tilde{h}}=\Omega^3 \sqrt{h}\,\\
\label{eqn:confn}
n_{\mu}\quad\rightarrow &\quad \tilde{n}_{\mu}=\Omega n_{\mu}=\Omega\left(-N,0\right),\\
\label{eqn:confnn}
n^{\mu}\quad\rightarrow &\quad \tilde{n}^{\mu}=\Omega^{-1} n^{\mu}=\Omega^{-1}\left(\frac{1}{N},\frac{-N^{i}}{N}\right),
\end{align}
where, by comparing \eqref{eqn:confn} with \eqref{eqn:confnn}, one can deduce that lapse and shift functions transform as
\begin{align}
\label{eqn:confN}
N\quad\rightarrow &\quad\,\,\tilde{N}=\Omega N,\\
\label{eqn:confNi}
N^{i}\quad\rightarrow &\quad\tilde{N}^{i}=N^{i}\quad  {\rm and}\quad \tilde{N}_{i}=\tilde{h}_{ij}\tilde{N}^{j}=\Omega^2 N_{i}.
\end{align}

 From the definition of extrinsic curvature \eqref{eqn:Kijdef}, it can
 be seen that it transforms nontrivially, 
\begin{equation}
\label{eqn:confK}
K_{ij}\quad\rightarrow\quad \tilde{K}_{ij}=\Omega K_{ij}+\frac{1}{\Omega}h_{ij}\mathcal{L}_{n}\Omega,
\end{equation}
which is due to the inhomogeneous transformation of its trace,
\begin{equation}
\label{eqn:confKtr}
K\quad\rightarrow\quad \tilde{K}=\frac{1}{\Omega}\left( K+3\mathcal{L}_{n}\Omega \right),
\end{equation}
while the traceless component transforms covariantly,
\begin{equation}
\label{eqn:confKt}
K_{ij}^{\ssst\rm T}\quad\rightarrow\quad \tilde{K}_{ij}^{\ssst\rm
  T}=\Omega K_{ij}^{\ssst\rm T}. 
\end{equation}
The infinitesimal conformal variation $\delta_{\omega}$ of $K_{ij}$
and its trace $K$ are then, respectively, given by 
\begin{align}
\label{eqn:infconfK}
\delta_{\omega}K_{ij}&=\omega K_{ij}+h_{ij}\mathcal{L}_{n}\omega\,,\\
\delta_{\omega}K&=-\omega K+3\mathcal{L}_{n}\omega\,.
\end{align}

 It follows from \eqref{eqn:conf3det} that the numerical power of $\Omega$,
 that is, the conformal weight, corresponds to the same power of
 ${(\sqrt{h})}^{1/3}$ in each of the objects
 \eqref{eqn:conf3met}--\eqref{eqn:confnn}. Therefore, if one rescales
 these objects by the appropriate power of ${(\sqrt{h})}^{1/3}$, one
 ends up with zero conformal weight objects, 
\begin{align}
\label{eqn:confinvobj}
\bar{N}^{i}&=N^{i}\,,\\
\bar{N}_{i}&={(\sqrt{h})}^{-2/3} N_{i}\,,\qquad\bar{N}={(\sqrt{h})}^{-1/3}N\,,\\
\bar{h}_{ij}&={(\sqrt{h})}^{-2/3}h_{ij}\,,\\
\bar{K}_{ij}^{\ssst\rm T}&={(\sqrt{h})}^{-1/3}K_{ij}^{\ssst\rm T}\,,
\end{align}
which are then all conformally invariant, except the scale
${(\sqrt{h})}^{1/3}$ itself (which is of conformal weight 1), and the
rescaled trace of the extrinsic curvature $K$ (of zero conformal
weight), 
\begin{equation}
\label{eqn:confobjK}
\bar{K}=\frac{1}{3}{(\sqrt{h})}^{1/3}K,
\end{equation}
because the scale is essential in their definition;
see \eqref{eqn:Ktrdef}. Thus, the only nonvanishing infinitesimal
conformal variations $\delta_{\omega}$ of the new variables are
$\delta_{\omega}\sqrt{h}$ and $\delta_{\omega}\bar{K}$, 
\begin{equation}
\label{eqn:infconf}
\delta_{\omega}{(\sqrt{h})}^{1/3}=\omega{(\sqrt{h})}^{1/3}\,,\quad
\delta_{\omega}\bar{K}=\bar{n}^{\mu}\del_{\mu}\omega. 
\end{equation} 

An important conclusion from here is that if a Lagrangian does not depend on
velocities $\dot{\sqrt{h}}$ and $\dot{K}$, the theory is conformally
invariant in the metric sector. Therefore, the tensor
densities \eqref{eqn:confinvobj}--\eqref{eqn:confobjK} are
natural choices for canonical variables in theories in which conformal
invariance or its breaking is of interest. Introducing $a:=
{(\sqrt{h})}^{1/3}$, we end up with
\eqref{eqn:VarsNbar}--\eqref{eqn:VarsKbar}. 

If matter is present, it is appropriate to introduce new, rescaled matter fields by
\begin{equation}
\label{eqn:cfinvphi}
\chi_{\ssst A} = a^n \varphi_{\ssst A}\,,
\end{equation}
such that $\chi$ does not transform under conformal transformations if
the conformal weight of the matter field $\varphi$ is $-n$. This has
already proven to be a useful substitution on many occasions in
the literature, for example in the use of Mukhanov-Sasaki variables
 for $n=1$; see, for example, \cite{BKK16}. The meaning of such a substitution
is the separation of the gravitational degrees of freedom from the pure
matter degree of freedom. It follows that the conformal transformation of
any field originates in the rescaling of the gravitational scale degree of
freedom only. 

\section{Unimodular decomposition of the connection and the
  $d$-dimensional Ricci tensor} 
\label{AppRicciBar}

We also need to know how to decompose the covariant derivative,
because the Christoffel symbols also decompose. One expects that the
covariant derivatives within the definitions of the extrinsic
curvature variables contain scale parts which are solely responsible
for the conformal transformation. Indeed, Christoffel symbols can be
decomposed into a part $\bar{\Gamma}^{k}_{\,\,\,ij}$ which is
determined solely by the unimodular part of the metric, and a part
$\sigma^{k}_{\,\,\,ij}$ which is solely responsible for the conformal 
transformation of the connection: 
\begin{align}
\label{eqn:GammaDec}
\Gamma^{k}_{\,\,\,ij}&=\bar{\Gamma}^{k}_{\,\,\,ij}+\sigma^{k}_{\,\,\,ij}\,,\quad {\rm where}\\
\label{eqn:Gammabar}
\bar{\Gamma}^{k}_{\,\,\,ij}&=\frac{1}{2}\bar{h}^{kl}\left(\del_{i}\bar{h}_{lj}+\del_{j}\bar{h}_{li}-\del_{l}\bar{h}_{ij}\right)\,,\\
\label{eqn:Cloga}
\sigma^{k}_{\,\,\,ij}&=\left(2\delta_{(i}^{k}\delta_{j)}^{l}-\bar{h}_{ij}\bar{h}^{kl}\right)\del_{l}\log a .
\end{align}
The individual parts have the properties
\begin{align}
\label{eqn:Gammabarprop}
\bar{\Gamma}^{i}_{\,\,\,ij}&=0\,,& \bar{h}^{ij}\bar{\Gamma}^{k}_{\,\,\,ij}&=-\del_{i}\bar{h}^{ik},\\
\label{eqn:Clogaprop}
\sigma^{i}_{\,\,\,ij}&=3\,\del_{j}\log
                  a\,,\quad&\bar{h}^{ij}\sigma^{k}_{\,\,\,ij}&=-\bar{h}^{kl}\del_{l}\log
                                                          a, 
\end{align}
because $h^{ij}\del_{k}\bar{h}_{ij}=0$.
Thus, the conformal transformation of the connection is solely due to
its scale part \eqref{eqn:Cloga}, which transforms as 
\begin{equation}
\label{eqn:ClogaConf}
\sigma^{k}_{\,\,\,ij}\rightarrow \tilde{\sigma}^{k}_{\,\,\,ij}=\sigma^{k}_{\,\,\,ij}+\left(2\delta_{(i}^{k}\delta_{j)}^{l}-\bar{h}_{ij}\bar{h}^{kl}\right)\del_{l}\log \Omega \,.
\end{equation}
 We call the quantity $\bar{\Gamma}^{k}_{\,\,\,ij}$ we call
 \textit{the conformal Riemannian connection}, or simply
 \textit{conformal Christoffel symbol}, because it is invariant under
 conformal transformations of the metric. The introduction of the
 conformal connection dates back to Thomas \cite{Thom1, Thom2}, who
 investigated conformal invariants and derived the Weyl and
 Schouten tensors using the transformation properties of the conformal
 connection. 

Let us now address the behavior of the covariant derivative of a
covariant vector density $V_{i}$ of weight $w$ under the unimodular
decomposition; we have 
\begin{align}
\label{eqn:covderdecom}
D_{i}V_{j}&=\bar{D}_{i}V_{j}-\sigma^{k}_{\,\,\,ij}V_{k}-w\,\del_{i}\log \sqrt{h} \,V_{j}\quad {\rm or}\\
\label{eqn:covderdecom1}
&=\bar{D}_{i}V_{j}-\sigma^{k}_{\,\,\,ij}V_{k}-w_{a}\,\del_{i}\log a \,V_{j}.
\end{align}
The term $\bar{D}_{i}V_{j}$ above will be explained shortly. In
\eqref{eqn:covderdecom1}, we have introduced $w_{a}:= 3w$, which we call
``the scale weight'', that is, the weight with respect to the scale density
$a$. For example, the unimodular part of the three-metric $\bar{h}_{ij}$
is a density of weight $w=-2/3$, or a density of scale weight
$w_{a}=-2$. In general, if an object has a conformal weight $m$, then
the scale weight of its conformally rescaled part is just its negative
[compare \eqref{eqn:VarsNbar}--\eqref{eqn:VarsKbar} with
\eqref{eqn:conf3met}--\eqref{eqn:confKtr}]. Due to \eqref{eqn:Cloga},
the second term in
\eqref{eqn:covderdecom1} is proportional to
$\del_{j}\log a$, and together with the third term (which is absent for
absolute tensors) yields the only scale part of the connection 
transforming under conformal transformations.  

The overbar on the covariant derivative in the first terms in
\eqref{eqn:covderdecom} and \eqref{eqn:covderdecom1} indicates that it
is completely determined by the unimodular part of the three-metric only,
that is, it is given in terms of \eqref{eqn:Gammabar} by 
\begin{equation}
\label{eqn:covderbar}
\bar{D}_{i}V_{j}:= \del_{i}V_{j}-\bar{\Gamma}^{k}_{ij}V_{k}\,.
\end{equation}
This could be called ``conformal covariant derivative'', but in fact,
the resulting quantity does not seem to be a tensor in general; hence
we say it is the ``scaleless part'' of the covariant derivative, or
the ``conformal part'' of the covariant derivative. We will call it
bar derivative, for short. 
In a similar way, we can define these bar derivatives on a tensor density of a general rank and weight---since the scale part does not enter its definition, there is no difference between cases for tensors of the same rank but different weight. For example, for rank-2 tensors we have
\begin{align}
\label{eqn:covderbar2}
\bar{D}_{k}T^{ij}&=\del_{k}T^{ij}+\bar{\Gamma}^{i}_{kl}T^{lj}+\bar{\Gamma}^{j}_{kl}T^{li}\,,\\
\label{eqn:covderbar11}
\bar{D}_{k}{T^{i}}_{j}&=\del_{k}{T^{i}}_{j}+\bar{\Gamma}^{i}_{kl}T^{l}_{j}-\bar{\Gamma}^{l}_{kj}{T^{i}}_{l}\,.
\end{align}
Note that the bar derivative of an arbitrary tensor density is not, in general,
scale independent and thus not conformally invariant. However, an
interesting case of \eqref{eqn:covderdecom} is its symmetrized
traceless part, for a vector density of the scale weight $w_{a}=-2$,
which turns out to be independent of the scale density, 
\begin{equation}
\left[D_{(i}V_{j)}\right]^{\ssst\rm T}\equiv\mathbb{1}_{ij}^{ab\ssst\rm T}D_{(i}V_{j)}=\left[\bar{D}_{(i}V_{j)}\right]^{\ssst\rm T}\,,
\end{equation}
and is thus conformally invariant. Such is the case with the second
term in \eqref{eqn:Kbardef}, leading to \eqref{eqn:Kbardef1}, which
makes $\bar{K}_{ij}^{\ssst\rm T}$ conformally invariant. Note in
passing that it can be easily shown that the above identity is valid
in any dimension $d$ if the vector density is of weight $w=-2/d$. 

Using \eqref{eqn:Clogaprop}, the covariant gradient of the
contravariant vector is simply given by 
\begin{equation} 
\label{eqn:covgradbar}
D_{i}V^{i}=\del_{i}V^{i}+\left(3-w_{a}\right)\del_{j}\log a \,V^{j}\,.
\end{equation}
Note that the factor of $3$ comes from the conversion
$\del_{i}\log\sqrt{h}=3(\del_{i}\log a)$. In the special case of
$w_{a}=3$, corresponding to a vector density of weight $w=1$, one
obtains the well-known formula $D_{i}V^{i}=\del_{i}V^{i}$. 
Because of \eqref{eqn:Gammabarprop}, we have in this
case $D_{i}V^{i}=\bar{D}_{i}V^{i}$.
Similarly, for tensor densities of higher
rank, the following expressions follow from \eqref{eqn:covderbar2} and
\eqref{eqn:covderbar11}: 
\begin{align}
\label{eqn:covgrad2}
\bar{D}_{i}T^{ij}&=\del_{i}T^{ij}+\bar{\Gamma}^{j}_{ik}T^{ik}\,,\\
\label{eqn:covgrad11}
\bar{D}_{i}{T^{i}}_{j}&=\del_{i}{T^{i}}_{j}-\bar{\Gamma}^{j}_{ik}{T^{i}}_{k}\,.
\end{align}
The difference between the barred covariant derivative and the usual
covariant derivative consists of all the scaleful terms arising from
the connection. Application to tensor densities of a general rank is
straightforward. One should, however, be aware of an exception, which
is $\bar{D}_{i}a$. This expression does not vanish, unlike its ``full
metric'' analog; this is because $\bar{D}$ is insensitive to scale density and does
not induce the weight-dependent term of the covariant derivative,
therefore leaving $\bar{D}_{i}a=\del_{i}a$. Additionally, we have
$\bar{D}_{i}\bar{g}_{kl}=0$, which follows easily from
$D_{i}g_{kl}=0$. In summary, the barred covariant derivative does not
recognize the difference between absolute tensors and tensor densities
of general weight. 

We are now equipped with all we need to decompose the Ricci
tensor. We do this for general dimension $d$, pretending that
indices and labels are not restricted to the three-dimensional case 
discussed in this paper [from now on, Latin indices 
assume values $i,j = 0,1,2,3... d-1$, and scale density is defined as
$a=(\sqrt{h})^{1/d}$]. Using \eqref{eqn:GammaDec} and
\eqref{eqn:Gammabarprop}, we have for the Ricci tensor 
\begin{align}
\label{eqn:Riccidefdec}
R_{ij}&=2\,\del_{[k}\Gamma^{k}_{\,\,\,i]j}+2\Gamma^{l}_{i[j}\Gamma^{k}_{\,\,\,k]l}\nonumber\\
&=\bar{R}_{ij}+2\,\del_{[k}\sigma^{k}_{\,\,\,i]j}+2\sigma^{l}_{i[j}\sigma^{k}_{\,\,\,k]l}-2\bar{\Gamma}^{k}_{l(i}\sigma^{l}_{\,\,\,j)k}+\bar{\Gamma}^{l}_{ij}\sigma^{k}_{\,\,\,kl},  
\end{align} 
where
\begin{equation}
\label{eqn:Riccibar}
\bar{R}_{ij}=\del_{k}\bar{\Gamma}^{k}_{\,\,\,ij}-\bar{\Gamma}^{l}_{ik}\bar{\Gamma}^{k}_{\,\,\,lj} 
\end{equation}
is the part of the Ricci tensor that does not depend on the scale density at
all; hence, it is conformally invariant.\footnote{In four dimensions,
  this is the so-called ``November tensor'' \cite{november} that
  Einstein proposed for his field 
  equations on his path towards the formulation of general
  relativity.} Explicitly expressed in terms of scale, the Ricci
tensor is 
\begin{align}
\label{eqn:Riccidecfin}
R_{ij}&=\bar{R}_{ij}-\left(d-2\right)\left(\delta_{(i}^{b}\delta_{j)}^{c}+\frac{1}{d-2}\bar{h}_{ij}\bar{h}^{bc}\right)\bar{D}_{b}\del_{c}\log a \nonumber\\
&\quad +\left(d-2\right)\left(\delta_{(i}^{b}\delta_{j)}^{c}-\bar{h}_{ij}\bar{h}^{bc}\right)\del_{b}\log a\,\del_{c}\log a\,.
\end{align}
The Ricci scalar is obtained by contracting the above expression with
$h^{ij}$; we, however, contract it with $\bar{h}^{ij}$, which corresponds to a rescaled Ricci scalar $a^2 R$, namely,
\begin{align}
\label{eqn:RicciScalDec}
& a^2 h^{ij}R_{ij}=\bar{h}^{ij}R_{ij}\nonumber\\
&=\bar{R}-2(d-1)\bar{h}^{ij}\biggl[\bar{D}_{i}\del_{j}\log a+\frac{(d-2)}{2}\del_{i}\log a \,\del_{j}\log a\biggr]\\[6pt]
\label{eqn:RicciScalDec1}
&=\bar{R}-\frac{2(d-1)}{a^2}\biggl[a\,\del_{i}\left(\bar{h}^{ij}\del_{j}a\right)+\frac{(d-4)}{2}\bar{h}^{ij}\del_{i}a\,\del_{j}a\biggr],
\end{align}
where $\bar{R}\equiv \bar{h}^{ij}\bar{R}_{ij}$ is conformally
invariant. The expression in the last line above is obtained by
employing $\del_{i}\del_{j}a = a\del_{i}\del_{j}\log a +a\del_{i}\log
a\,\del_{j}\log a$. Therefore, only the second and third terms
transform under conformal transformations, producing the following
expression: 
\begin{align}
\label{eqn:Rbarconf}
\Delta_{\Omega}&\left(a^2 R\right):=\tilde{a}^2\tilde{R}-a^2 R\nonumber\\
&=-2\left(d-1\right)\Biggl(D_{i}\left(\bar{h}^{ij}\del_{j}\log \Omega\right)\nonumber\\
&\qquad\qquad+\frac{1}{2}\left(d-2\right)\bar{h}^{ij}\del_{i}\log\Omega\,\del_{j}\log\Omega\Biggr)\\ 
&=-\frac{2\left(d-1\right)}{ \Omega^2}\Biggl(\Omega
  D_{i}\left(\bar{h}^{ij}\del_{j}
  \Omega\right)+\frac{\left(d-4\right)}{2}\bar{h}^{ij}\del_{i}\Omega\,\del_{j}\Omega\Biggr), 
\end{align}
which turns out to be exactly the expected well-known difference after
a conformal transformation of the rescaled Ricci scalar. A rule of
thumb can be used to quickly determine the conformal transformation of
an object in question: simply make a substitution
$\bar{D}_{i}\rightarrow D_{i}$ in the first term in
\eqref{eqn:RicciScalDec} or $\del_{i}\rightarrow D_{i}$ in the first
term in \eqref{eqn:RicciScalDec1}, along with $a\rightarrow \Omega
$. 

It is useful to take a look at the traceless Ricci tensor, since it is
of importance in the W theory. It turns out to be 
\begin{align}
\label{eqn:RicciTless}
R_{ij}^{\ssst\rm T}&=R_{ij}-\frac{1}{d}h_{ij}R=\bar{R}_{ij}-\frac{1}{d}\bar{h}_{ij}\bar{R}\\[6pt]
&=\bar{R}_{ij}^{\ssst\rm T}-\left(d-2\right)\left(\bar{D}_{i}\del_{j}\log a-\del_{i}\log a\,\del_{j}\log a\right)^{\ssst\rm T},
\end{align}
where, again, ``$\ssst \rm T$'' denotes that the trace has been projected away from an object by contraction with traceless identity \eqref{eqn:oneT}.

Note that one could define a quantity ${\bar{R}^{i}}_{\,\,\,jkl}$
which would be the part of the Riemann tensor determined only by the
conformal connection ${\bar{\Gamma}}^{k}_{\,\,\,ij}$ \cite{Thom1}, in
analogy to $\bar{R}_{ij}$. It is expected that the Weyl tensor---due
to its conformal invariance---could be determined from
$\bar{h}_{ij}$ only as well, but what is its relationship with
${\bar{R}^{i}}_{\,\,\,jkl}$? One may then seek such a relationship
from the decomposition of the Riemann tensor. Namely, the irreducible
decomposition of the Riemann tensor under the Lorentz group is 
\begin{equation}
\label{eqn:RiemWeyldec}
 {R^{i}}_{jkl}={C^{i}}_{jkl}+\left(\delta^{i}_{[k}R_{l]j}-g_{j[k}{R^{i}}_{l]}\right)+\frac{1}{3}\delta_{[l}^{i}g_{k]j}R\,,
\end{equation}
but it can also be written in terms of the Schouten tensor $S_{ij}$,
\begin{equation}
\label{eqn:RWSchdec}
  {R^{i}}_{jkl}={C^{i}}_{jkl}+\frac{2}{d-2}\left(\delta^{i}_{[k}S_{l]j}-g_{k[j}S^{i}_{l]}\right)\,,
\end{equation}
where the Schouten tensor is defined by
\begin{align}
\label{eqn:Schouten}
S_{ij}&=R_{ij}-\frac{1}{2\left(d-1\right)} h_{ij}R.
\end{align}
Expressing now the Weyl tensor from \eqref{eqn:RiemWeyldec},
\begin{equation}
\label{eqn:WeyldefSch}
 {C^{i}}_{jkl}={R^{i}}_{jkl}-\frac{2}{d-2}\left(\delta^{i}_{[k}S_{l]j}-g_{k[j}S^{i}_{l]}\right)
\end{equation}
and recalling its independence on scale, a separate scale dependence from
the two terms on the right-hand side should cancel. 
 Therefore, it should be possible to define the Weyl tensor
equivalently in terms of the barred version of the quantities on
the right-hand side only. Indeed, by defining ${R^{i}}_{jkl}$ as
explained at the beginning of this paragraph, with decomposing the
Schouten tensor into its conformally invariant part [$\bar{S}_{ij}$,
as the barred version of \eqref{eqn:Schouten}] and scale part
$W_{ij}$, 
\begin{align}
\label{eqn:Sch-dec}
S_{ij}&=\bar{S}_{ij}+W_{ij}\,,\\
\label{eqn:Sch-bar}
\bar{S}_{ij}&:=\bar{R}_{ij}-\frac{1}{2\left(d-1\right)} \bar{h}_{ij}\bar{R}\\
\label{eqn:Sch-scale}
W_{ij}&:=-(d-2)\Bigg[\bar{D}_{i}\del_{j}\log a\nonumber\\
&-\left(\mathbb{1}_{(ij)}^{bc}-\frac{1}{2}\bar{h}_{ij}\bar{h}^{bc}\right)\del_{b}\log a\,\del_{c}\log a\Bigg],
\end{align}
one arrives at the same definition of the Weyl tensor obtained (using
transformation properties of the conformal connection and eliminating
the scale dependent terms from its antisymmetrized derivatives) by
Thomas \cite{Thom2}, 
\begin{equation}
\label{eqn:WeyldefBar}
 {C^{i}}_{jkl}={\bar{R}^{i}}_{\,\,\,jkl}-\frac{2}{d-2}\left(\delta^{i}_{[k}\bar{S}_{l]j}-g_{k[j}{\bar{S}^{i}}_{l]}\right).
\end{equation}
Note that both terms on the right-hand side are separately conformally
invariant, but are not tensorial quantities. Therefore, the Weyl
tensor may be interpreted as the traceless part of the unimodular part
${\bar{R}^{i}}_{\,\,\,jkl}$ of the Riemann tensor. One concludes that
the conformal transformation of the Riemann tensor arises solely from
the expression 
\begin{equation}
{W^{i}}_{jkl}:=\frac{2}{d-2}\left(\delta^{i}_{[k}W_{l]j}-g_{k[j}W^{i}_{l]}\right)\,,
\end{equation}
where $W_{ij}$ is given by \eqref{eqn:Sch-scale}.

\section{Various identities}
\label{AppIdent}

Starting from the split of the velocity $\mathcal{L}_{n}K_{ij}$ into
its traceless and trace part, 
\begin{equation}
\label{eqn:LieKsplit}
\mathcal{L}_{n}K_{ij}=\left(\mathcal{L}_{n}K_{ij}\right)^{\ssst\rm T}+\frac{1}{3}h_{ij}h^{ab}\mathcal{L}_{n}K_{ab}\,,
\end{equation}
using the traceless-trace decomposition of the extrinsic curvature
\eqref{eqn:Ksplit} as well as the following idendity, 
\begin{equation}
\label{eqn:LieKtr}
\mathcal{L}_{n}K=h^{ab}\mathcal{L}_{n}K_{ab}-2K_{ab}K^{ab},
\end{equation}
where $K^{ab}=-\mathcal{L}_{n}h_{ab}/2$, one can show that the traceless part of $\mathcal{L}_{n}K_{ij}$ can be expressed in terms of the Lie derivative of $K_{ij}^{\ssst\rm T}$,
\begin{equation}
\label{eqn:tLieKt}
\left(\mathcal{L}_{n}K_{ij}\right)^{\ssst\rm T}=\mathcal{L}_{n}K^{\ssst\rm T}_{ij}+\frac{2}{3}K_{ij}^{\ssst\rm T}K-\frac{2}{3}h_{ij}K_{ab}^{\ssst\rm T}K^{ab\ssst\rm T}.
\end{equation}
This expression is then used to rewrite \eqref{eqn:elW} in terms of the Lie derivative of $K_{ij}^{\ssst\rm T}$, namely,
\begin{equation}
\label{eqn:tLieKtelW}
\left(\mathcal{L}_{n}K_{ab}\right)^{\ssst\rm T}-K_{ab}^{\ssst \rm T}K=\mathcal{L}_{n}K^{\ssst\rm T}_{ij}-\frac{1}{3}K_{ij}^{\ssst\rm T}K-\frac{2}{3}h_{ij}K_{ab}^{\ssst\rm T}K^{ab\ssst\rm T}\,,
\end{equation}
which is then used with unimodular-conformal variables in Sec. \ref{ClassW}, to arrive at \eqref{eqn:elWsmpl}. Also note that by taking the trace of the above equation, we obtain
\begin{equation}
h^{ij}\mathcal{L}_{n}K^{\ssst\rm T}_{ij}=2K_{ij}^{\ssst\rm T}K^{ij\ssst\rm T}\,,
\end{equation}
which actually simply follows also from $\mathcal{L}_{n}h^{ij}K_{ij}^{\ssst\rm T}=0$.

The appearance of $R_{ij}$ and covariant derivatives in an expression
introduces a dependence on the scale density $a$, but there is a particular type of
operator that can be constructed with these two objects such that this
dependence vanishes when they are contracted with some particular
tensor densities. Such operators are met in many places in physics,
and it appears in the present paper, too. The motivation for seeking
such an operator in general can be found in the expression for
$C_{ij}^{\ssst\rm T}$ given by \eqref{eqn:elWsmpl}. Since the square of
$C_{ij}^{\ssst\rm T}$ is the kinetic term (of the electric part) of
the Weyl action, it is expected that it is itself conformally
invariant, that is, scaleless and independent of $\bar{K}$. Now, all terms except
$\,^{\ssst (3)}\! R_{ij}^{\ssst\rm T}$ and $(D_{j}\del_{j}N)^{\ssst\rm
  T}/N$ are manifestly scaleless (and $\bar{K}$-less). To show that
$\,^{\ssst (3)}\! R_{ij}^{\ssst\rm T} + (D_{j}\del_{j}N)^{\ssst\rm
  T}/N$ is conformally invariant, we use the unimodular-conformal
decomposition to prove that the scale density $a$ indeed vanishes from this
expression, and the proof will be given for any dimension [again, $i,j =
0,1,2,3... d-1$, and the scale density is defined as $a=(\sqrt{h})^{1/d}$]. We
also introduce a conformally invariant scalar density
$\bar{\phi}=a^{-1}\phi$, where $\phi$ is a general scalar of conformal
weight $-1$, in order to generalize the case with the lapse
function. We do not go here into the question of whether or not the
result is valid for a tensor of arbitrary rank and density, but leave
this for another place. 

Let us take a look at the two objects $\,^{\ssst (3)}\! R_{ij}$ and $(D_{i}\del_{j}\phi)/\phi$ separately. We already have $\,^{\ssst (3)}\! R_{ij}$ in \eqref{eqn:Riccidecfin}. Then,
\begin{align}
\label{eqn:DDphi}
\frac{D_{i}\del_{j}\phi}{\phi}&=\frac{D_{i}D_{j}\bar{\phi}}{\bar{\phi}}\nonumber\\
&=\frac{1}{\bar{\phi}}\bar{D}_{i}\del_{j}\bar{\phi}+\bar{D}_{i}\del_{j}\log a+\bar{h}_{ij}\bar{h}^{bc}\del_{b}\log a\,\del_{c}\bar{\phi}\nonumber\\
&\quad-\left(2\delta_{(i}^{b}\delta_{j)}^{c}-\bar{h}_{ij}\bar{h}^{bc}\right)\del_{b}\log a\,\del_{c}\log a \,.
\end{align}
Now observe that the second and fourth terms are appearing (up to
$d$-dependent coefficients) in $R_{ij}$ in \eqref{eqn:Riccidecfin}
with an opposite sign. However, the third term contains
$\del_{c}\bar{\phi}$ and cannot be found in there, and is present without an opposite-signed pair in $R_{ij}$ to be canceled with. But this term is in its totality a part of the trace of expression \eqref{eqn:DDphi},
\begin{align}
\bar{h}^{ij}\frac{D_{i}D_{j}\bar{\phi}}{\bar{\phi}}&=\frac{1}{\bar{\phi}}\bar{h}^{ij}\bar{D}_{i}\del_{j}\bar{\phi}+\bar{h}^{ij}\bar{D}_{i}\del_{j}\log a+d\,\bar{h}^{bc}\del_{b}\log a\,\del_{c}\bar{\phi}\nonumber\\
&\quad-\left(d-2\right)\bar{h}^{bc}\del_{b}\log a\,\del_{c}\log a \,,
\end{align}
which means that the \textit{traceless} part of \eqref{eqn:DDphi} does not contain it. Therefore, this ``coincidence'' can be used to form a traceless operator from \textit{traceless parts} $\,^{\ssst (3)}\! R_{ij}^{\ssst\rm T}$ and $(D_{i}\del_{j}\phi)^{\ssst\rm T}/\phi$,
\begin{align}
\label{eqn:DDRT}
R_{ij}^{\ssst\rm T}&+\left(d-2\right)\frac{1}{\phi}D_{ij}^{\ssst\rm T}\phi \nonumber\\
&= \bar{R}_{ij}^{\ssst\rm T}-\left(d-2\right)\left(\bar{D}_{i}\del_{j}\log a-\del_{i}\log a\,\del_{j}\log a\right)^{\ssst\rm T}\nonumber\\
&+ \left(d-2\right)\left(\frac{1}{\bar{\phi}}\bar{D}_{i}\del_{j}\bar{\phi}+\bar{D}_{i}\del_{j}\log a -\del_{i}\log a\,\del_{j}\log a\right)^{\ssst\rm T}\nonumber\\
&= \bar{R}_{ij}^{\ssst\rm T}+\left(d-2\right)\frac{1}{\bar{\phi}}\left[\bar{D}_{i}\del_{j}\bar{\phi}\right]^{\ssst\rm T},
\end{align}
which is indeed manifestly conformally invariant. Let us now write this operator in a more recognizable form, by dividing it with $(d-2)$ and switching the order of terms,
\begin{align}
\label{eqn:DDRTeq}
\left((D_{i}D_{j})^{\ssst\rm T}+\frac{1}{d-2}R_{ij}^{\ssst\rm T}\right)\phi= \left(\bar{D}_{i}\del_{j}+\frac{1}{d-2}\bar{R}_{ij}\right)^{\ssst\rm T}\phi\,.
\end{align}
The same operator appears in \eqref{eqn:elWsmpl}, with $d=3$. Namely,
the traceless momentum density $\bar{P}^{ij}$ of scale weight
$\omega_{a}=4$, contracted with $\left((D_{i}D_{j})^{\ssst\rm
    T}+\frac{1}{d-2}R_{ij}^{\ssst\rm T}\right)$ gives 
\begin{align}
\label{eqn:DDRTPeq}
\left((D_{i}D_{j})^{\ssst\rm T}+\frac{1}{d-2}R_{ij}^{\ssst\rm T}\right)\bar{P}^{ij}\stackrel{d=3}{=} \left(\del_{i}\bar{D}_{j}+\bar{R}_{ij}\right)^{\ssst\rm T}\bar{P}^{ij}\,,
\end{align}
which matches the expression in \eqref{eqn:HamWcf}. These derivations
add to the power of the method of using the unimodular-conformal decomposition.

We finally make the interesting observation that the very same operator
considered above is precisely the one that appears in the Bach
equations, which are conformally invariant. Setting $d=4$, contraction
of $\left(D_{i}D_{j}+\frac{1}{d-2}R_{ij}\right)$ with the Weyl tensor
ensures that the operator is traceless, thus eliminating all the
scale-dependent terms from it. That is why the Bach equations
\eqref{eqn:Bach} can be simplified to 
\begin{equation}
\label{eqn:BachSmpl}
\left(\nabla_{k}\nabla_{l}+\frac{1}{2}R_{kl}\right)C^{k\,\,\,l}_{\,\,\,i\,\,\,j}=
\left(\bar{\nabla}_{k}\bar{\nabla}_{l}+\frac{1}{2}\bar{R}_{kl}\right)C^{k\,\,\,l}_{\,\,\,i\,\,\,j}=0\,,
\end{equation}
which is manifestly scaleless and thus conformally invariant.

However, a question of generality poses itself. The above examples
refer to a scalar density of scale weight $\omega_{a}=-1$,
\eqref{eqn:DDRT}, a traceless rank 2 tensor density of scalar weight
4, \eqref{eqn:DDRTPeq}, and the Weyl tensor, which is a rank 4
traceless tensor, but is also conformally invariant,
\eqref{eqn:BachSmpl}. One could, in principle, derive a general
expression for the operator
$\left(D_{i}D_{j}+\frac{1}{d-2}R_{ij}\right)^{\ssst\rm T}$ contracted
with a tensor of arbitrary rank, symmetry and density, in arbitrary
dimensions, such that the resulting expression is conformally
invariant, but that question deserves a separate study. 


\section*{Acknowledgments}
We are grateful to Brian Pitts, Tatyana Shestakova, and Alexei
Starobinsky for interesting discussions.



\end{document}